\documentclass[reprint,prd,amsmath,amssymb,aps,lengthcheck,floatfix,twocolumn,superscriptaddress]{revtex4-1}
\usepackage{balance}
\usepackage{hyperref}
\usepackage{graphicx}
\usepackage{amsfonts}
\usepackage{amssymb}
\usepackage{dcolumn}
\usepackage{bm}
\usepackage{ulem}
\usepackage[utf8]{inputenc}
\usepackage[T1]{fontenc}
\usepackage{amsmath}
\usepackage[version=4]{mhchem}
\usepackage{stmaryrd}
\usepackage{xcolor}
\usepackage{color}
\usepackage{subeqnarray}
\newcommand{\bwt}{\begin{widetext}}
\newcommand{\ewt}{\end{widetext}}
\newcommand{\beq}{\begin{equation}}
\newcommand{\eeq}{\end{equation}}
\newcommand{\bea}{\begin{eqnarray}}
\newcommand{\eea}{\end{eqnarray}}

\begin{document}
\title{Coupled nuclear and leptonic longitudinal collective modes in neutron star matter : a covariant Vlasov approach}
\author{Aziz Rabhi} 
\email{rabhi@uc.pt}
\affiliation{IPEST La Marsa, Carthage University, Tunisia.}
\affiliation{CFisUC, Department of Physics, University of Coimbra, 3004-516 Coimbra, Portugal.}
\author{Olfa Boukari} 
\email{olfa.boukari@isepbg.ucar.tn}
\affiliation{ISEP-BG La Soukra, Carthage University, Tunisia.}
\author{Sidney S. Avancini} 
\email{avancini@ufsc.br}
\affiliation{Departamento de F\'{\i}sica, Universidade Federal de Santa Catarina, Florian\'opolis, SC, CP. 476, CEP 88.040-900, Brazil.} 
\author{Constan\c ca Provid\^encia}
\email{cp@uc.pt}
\affiliation{CFisUC, Department of Physics, University of Coimbra, 3004-516 Coimbra, Portugal.}

\begin{abstract}
A covariant relativistic approach based on the Vlasov equation is used to study collective modes in neutron-star matter. The analysis is carried out within relativistic mean-field models describing charge-neutral and $\beta$-equilibrated matter composed of neutrons, protons, electrons, and muons. We investigate the conditions under which nuclear collective excitations couple to electron and muon plasmon modes, a phenomenon relevant for neutron stars and supernova matter. The study is undertaken considering relativistic mean field models with different isoscalar and isovector properties.  It is shown that the nuclear–leptonic coupling can be sufficiently strong to modify the onset of nuclear collective modes and to affect their isoscalar or isovector character.
\end{abstract}
\maketitle

\section{Introduction}
The study of compact astrophysical objects such as neutron stars and supernova cores requires a comprehensive theoretical framework that combines elements of nuclear physics, particle physics, astrophysics, and statistical mechanics. At densities ranging from sub-saturation to several times nuclear saturation density, the matter in these environments consists primarily of neutrons, protons, electrons, and, at higher densities, muons. Neutrinos may also be present if their mean free path is sufficiently short. Charge neutrality is maintained by electrons and muons, effectively canceling the Coulomb contribution from protons.

In addition to a precise determination of the equation of state (EOS) of dense matter, an accurate description of neutrino transport is essential for modeling the thermal and dynamical evolution of these systems. Neutrino opacity is significantly influenced by nucleon-nucleon interactions, particularly through coherent scattering off density fluctuations~\cite{sawyer1975, iwamoto1982}. Both single-particle and collective effects contribute to this process. As a result, understanding the collective modes in neutron star matter in $\beta$-equilibrium is essential for reliable predictions of neutrino propagation in dense astrophysical environments. Collective modes  have been discussed within a relativistic mean field approach in several works \cite{Chin:1977iz,Lim1989,Nielsen:1990ft,Nielsen:1993zz,greco2003,avancini05,cp2006, Brito2006} and, more recently, also in \cite{Ye:2023fhy,rabhi2025,Ye:2025ijq}.

Recently, we developed a covariant formulation of the Vlasov equation and applied it to several problems concerning relativistic nuclear matter. In particular, this framework was used to study the stability of neutron–proton–electron matter under strong magnetic fields~\cite{avancini2018}, and to analyze longitudinal collective modes in asymmetric magnetized nuclear matter~\cite{rabhi2026}.

In the  study~\cite{rabhi2025}, we employed the covariant relativistic approach based on the Vlasov equation to investigate collective longitudinal modes in infinite asymmetric nuclear matter within a set of relativistic mean-field (RMF) models with non-linear meson terms. Dispersion relations for isoscalar and isovector modes were derived over a wide range of densities, isospin asymmetries, and momentum transfers, employing 11 RMF parameterizations with various nuclear matter properties. Special attention was given to $\beta$-equilibrated matter. It was found that the propagation of isoscalar- and isovector-like collective modes is strongly influenced by the density dependence of both the symmetric nuclear matter equation of state (EOS) and the symmetry energy. In particular, a stiff EOS was shown to favor the propagation of isoscalar-like modes at high densities, as also discussed in \cite{Ye:2023fhy}, while a stiff symmetry energy enhanced the presence of isovector-like modes at densities up to approximately twice the saturation density. The coupling of nuclear modes to the electron plasmon was also examined.

In the present study, we extend this formalism to describe the collective modes in fully $\beta$-equilibrated neutron star matter, incorporating both nucleonic and leptonic degrees of freedom. This provides a more realistic description of the outer core of neutron stars and allows us to explore how the composition and EOS influence the spectrum of collective excitations at supra-saturation densities. Previous studies have shown the existence of an isovector-like mode at low densities, whose onset depends on isospin asymmetry, while at higher densities (two to three times saturation density) the mode evolves into an isoscalar-like character, suggesting a possible identification of a transition density \cite{greco2003,avancini05}. We further examine how the presence of electrons modifies these excitations, particularly for long-wavelength modes where protons and electrons move coherently.

Collective leptonic excitations (plasmons) have been widely studied in both nonrelativistic and relativistic frameworks, with relativistic treatments allowing for additional dissipation channels such as one-photon pair creation \cite{Jancovici1962,BraatenSegel1993}. Since neutrino emission in compact stars is sensitive to the dispersive properties of dense matter, such effects may play a relevant role in transport phenomena, particularly when pair creation becomes kinematically allowed and competes with neutrino-emission processes \cite{Yakovlev2001,Kantor2007,Jaikumar2005}.

In Sec.~\ref{sectII}, we briefly review the covariant Vlasov formalism and its extension from neutron-proton-electron (npe) matter, as developed in Ref.~\cite{rabhi2025}, to neutron-star matter composed of neutrons, protons, electrons, and muons (npe$\mu$ matter). Numerical results are presented and discussed in Sec. III. Finally, the main conclusions are summarized in Sec. IV.

%
%
%
%
%
%
%
%
%
\section{Covariant Vlasov equation formalism}
\label{sectII}
For the description of the EoS of neutron star matter, we employ a field-theoretical approach in which the nucleons interact via the exchange of $\sigma$-$\omega$-$\rho$ mesons.  
The Lagrangian density using natural units, i.e., taking $c=\hbar=$1, can be written as
\begin{equation}
{\cal L}=\sum_{j=p,n,e,\mu} {\cal L}_j + \cal L_\sigma + {\cal L}_\omega +
{\cal L}_\rho + {\cal L}_{\omega \rho } + {\cal L}_{A} \ ,
\label{lag}
\end{equation}
with
\begin{equation}
{\cal L}_j=\bar \psi^{(j)}\left[\gamma^\mu i D_\mu^{(j)}-M^*_j \right]\psi^{(j)} ,\nonumber
\end{equation}
where the Dirac field $\psi^{(j)}$ describes a single nucleon, a proton, or a
neutron, the covariant derivative is defined as $ iD^{(j)}_{\mu}~= ~ i \partial_{\mu} - 
{\cal V}^{(j)}_{\mu} $,
and the index $(j)=(n,p,e,\mu)$, stands for the neutron, proton, electron, and muon,
\begin{equation}
 {\cal V}^{(j)}_{\mu} = 
 \left\{
       \begin{array}{l}
         g_v \, V_\mu  + \frac{\tau_{j}}{2}\, g_\rho\, \vec{b}_\mu+ Q_{j}\, A_\mu \ ~ ,~j=p, n \\ 
         \\
         Q_j\, A_\mu \ ~ ,~ j=e, \mu
     \end{array}    \right. \ ,  \label{eq2}
\end{equation}
$M^*_p=M^*_n=M^*=M-g_s\phi(x),$ $M^{*}_{e}=m_{e}$, $M^{*}_{\mu}=m_{\mu}$,  with isospin values $\tau_p=+1$, $\tau_n=-1$ and electric charges $Q_p=+e$, $Q_e=Q_\mu=-e$.
$e=\sqrt{4\pi/137}$ is the electromagnetic coupling constant.  To investigate the impact of the density dependence of the nuclear symmetry energy on collective modes in cold neutron star matter, we employ three models consistent with constraints from multimessenger resonance shattering flares~\cite{Duncan2023}.
For the nuclear matter parameters, we will consider the following set of RMF models: FSU2H~\cite{tolos17,tolos17_2,Negreiros18}, NL3~\cite{nl3} and NL3~$\omega\rho$ \cite{Pais16, Horowitz01}.
The meson and photon contributions in the Lagrangian density, Eq.~(\ref{lag}) are given by
\begin{eqnarray}
\mathcal{L}_{{\sigma }} &=&\frac{1}{2}\left( \partial _{\mu }\phi \partial %
^{\mu }\phi -m_{s}^{2}\phi ^{2}-\frac{1}{3}\kappa \phi ^{3}-\frac{1}{12}%
\lambda \phi ^{4}\right) \ , \nonumber \\
\mathcal{L}_{{\omega }} &=&\frac{1}{2} \left(-\frac{1}{2} \Omega _{\mu \nu }
\Omega ^{\mu \nu }+ m_{v}^{2}V_{\mu }V^{\mu }
+\frac{1}{12}\xi g_{v}^{4}(V_{\mu}V^{\mu })^{2} 
\right) \ , \nonumber \\
\mathcal{L}_{{\rho }} &=&\frac{1}{2} \left(-\frac{1}{2}
{\vec{B}}_{\mu \nu }\cdot {\vec{B}}^{\mu
\nu }+ m_{\rho }^{2} \vec{b}_{\mu }\cdot \vec{b}^{\mu } \right)   \ , \nonumber \\
\mathcal{L}_{\omega \rho } &=& \Lambda_v g_v^2 g_\rho^2 V_{\mu }V^{\mu }
\vec{b}_{\nu }\cdot \vec{b}^{\nu }   \nonumber \\
\cal{L}_{A} &=&-\frac{1}{4} F_{\mu \nu }F ^{\mu \nu }~, \label{mesonlag}
\end{eqnarray}
where $\Omega _{\mu \nu }=\partial _{\mu }V_{\nu }-\partial _{\nu }V_{\mu }$, 
$\vec{B}_{\mu \nu }=\partial _{\mu }\vec{b}_{\nu }-\partial _{\nu }
\vec{b}
_{\mu }-\Gamma_{\rho }(\vec{b}_{\mu }\times \vec{b}_{\nu })$ and 
$F_{\mu \nu }=\partial _{\mu }A_{\nu }-\partial _{\nu }A_{\mu }$.
The parameters $\kappa$, $\lambda$, and $\xi$ denote the self-interaction couplings, while $\Lambda_v$ is the coupling of the $\omega$–$\rho$ mixing term, introduced to soften the density dependence of the symmetry energy above saturation density.
From the Euler–Lagrange equations, the Dirac equation governing the fermion fields can be derived:
\begin{eqnarray}
 i \gamma^\mu D^{(j)}_{\mu}~\psi^{(j)} = M^{\star}_{j} ~\psi^{(j)} \ , \label{dirac1}
\end{eqnarray}
and its conjugate equation:
\begin{eqnarray}
 \bar{\psi}^{(j)} i D^{\dagger (j)}_{\mu}\gamma^\mu ~= - M^{\star}_{j} ~\bar{\psi}^{(j)} \, 
 \label{dirac2}
\end{eqnarray}
where $ iD^{\dagger (j)}_{\mu} = ~ i\overleftarrow{\partial}_{\mu} + {\cal V}^{(j)}_{\mu} $.
In charge-neutral, $\beta$-equilibrated matter, the following conditions must be satisfied:
\begin{equation}
   \mu_n - \mu_p = \mu_{e} = \mu_{\mu} \, \quad \text{and} \, \quad   \rho_ p= \rho_e + \rho_\mu \,.          
     \label{eqbeta}
\end{equation}
In the following section, we extend our investigation from asymmetric npe matter~\cite{rabhi2025} to beta-equilibrated matter including electrons and muons through the derivation of the Vlasov equation for a hadronic system, starting from the general transport equations. Our formalism is based on the covariant Wigner function described in detail in Ref.~\cite{avancini2018, rabhi2025}. Here, we present only the main results relevant to transport theory. We focus on the new technical aspects that arise when applying the covariant Wigner function to beta-equilibrated matter. This formalism is well suited for studying collective modes and paves the way for investigating thermal and electrical conductivities in such matter.
\subsection{Covariant Vlasov approach}
We begin with the generalized Vlasov equation established in Ref.~\cite{rabhi2025}, extended here to include both electrons and muons, expressed as:
 \begin{equation}
\partial_t f_{(j)} + \vec{v} \cdot \nabla_x f_{(j)} +
  (\vec{{\cal E}} + \vec{v} \times \vec{\cal B})\cdot \nabla_p f_{(j)} = 0, \, j=p, n, e, \mu
  \label{vlasov1}
 \end{equation}
where $\vec{v}=\vec{p}/E_p$ and
\begin{eqnarray}
 && {\cal E}_i^{(j)} = {\cal F}_{0i}^{(j)} - \frac{M_j^\star(x)}{E_p^{(j)}} 
 \nabla_{x,i}~ M_j^\star(x) ~\ , \  j=p,n \nonumber \ , \\
 && {\cal E}_i^{(j)} =  {\cal F}_{0i}^{(j)},~\, j=e,\mu  \nonumber \ , \\
 && {\cal B}^{(j)}_i ~=~\epsilon_{ilm} \partial_{x,l} {\cal V}^{(j)}_m ~,~j=p,n,e,\mu~\,
\end{eqnarray}
$i,l,m=1,2,3$, $ {\cal F}_{0i}^{(j)}=- \partial_t ~\delta \vec{{\cal V}}^{(j)} - \nabla_x \delta {\cal V}_0^{(j)} $. Here, $f^{(j)}$ is the distribution function of each species.

To obtain the dispersion relations we analyze the current densities of baryons and leptons,
$$
\begin{aligned}
 J_{\mu}(x)&=\sum_{j=n,p,e}\frac{2}{(2\pi)^3}\int \frac{d^3 p}{p^0} p_{\mu} f^{(j)}(x, \vec{p}) \\
 & = \sum_{j=n,p,e,\mu} J_{\mu}^{(j)} (x), \quad p^{0}=E_{j}^{(0)}=\sqrt{\vec{p}^2+{M^{* (0)}_j}^2}.
\end{aligned}
$$
where $M_j^{\star (0)}= M-g_s \phi^{(0)}$, $M_e^{\star (0)}$=$m_e$ and $M_\mu^{\star (0)}$=$m_\mu$. The evolution of these currents is determined by the generalized Vlasov equation,~Eq.(\ref{vlasov1}), which provides the microscopic conservation properties of the system. Consequently, one obtains the following continuity equation: $\partial^{\mu}J^{(j)}_{\mu}=0$.

We will consider small oscillations in relation to the equilibrium generated by perturbations of the distribution function as
\begin{equation}
 f^{(j)}(\vec{p}) =f^{(0)(j)}(\vec{p})+\delta f^{(j)}(\vec{p}).
 \label{delphi}
\end{equation}
Since in the present work we are only interested in systems at zero temperature, the equilibrium distribution function is given by 
\beq
f^{(0)(j)}(\vec{p})=\theta\left(p^2_{Fj} -p^2 \right), j=p, n, e, \mu,
\label{wignereq5}
\eeq 
where the Heaviside function $\theta(x)$ was used.

The small perturbation of the distribution functions, $f^{(0)(j)}$, around their equilibrium
values, will generate perturbations on the fields:
\begin{eqnarray}
&& \phi = \phi^{(0)}+\delta \phi ~,~V_\mu=V_\mu^{(0)}+\delta V_\mu ~,~
 b_\mu=b_\mu^{(0)}+\delta b_\mu~, \nonumber \\
 &&~A_\mu=A_\mu^{(0)}+ \delta A_{\mu}~ , \label{smalldev}
\end{eqnarray}
and cause a corresponding perturbation of the equilibrium 4-current,
\begin{equation}
  J^{(j)}_\mu(x) = J^{(0)(j)}_\mu
   + \delta J^{(j)}_\mu \ ,
\end{equation}
with
\begin{equation}
  \delta J^{(j)}_\mu = \frac{2}{(2\pi)^3} \int \frac{d^3 p}{E^{(0)}_j}
    ~p_\mu ~\delta f^{(j)}. \label{curre}
\end{equation}
After substituting Eq.(\ref{delphi}) in the Vlasov equation, Eq.(\ref{vlasov1})
retaining only terms of the first order in $\delta f^{(j)}$, one obtains:
 \begin{eqnarray}
  && \partial_t \delta f^{(j)} + \vec{v} \cdot \nabla_x \delta f^{(j)} +
  \vec{v} \times ( \nabla_x \times \vec{\cal V}^{(0)(j)} ) \cdot \nabla_p \delta f^{(j)} 
  \cr
  &&
  + \left[ \vec{v} \times \nabla_x \times (\vec{\cal V}^{(0)(j)}  + 
\delta \vec{\cal V}^{(j)} ) + g_s \frac{{M_j^\star}^{(0)}}{{E^{(0)}_j}} \nabla_x \delta \phi 
\right.  \nonumber \\ &&  \left.   
- \partial_t ~\delta \vec{{\cal V}}^{(j)} -
 \nabla_x \delta {\cal V}_0^{(j)}  \right]  \cdot 
 \nabla_p f^{(0)(j)}    
   = 0 \ ,
  \label{dvlasov}
 \end{eqnarray}
where $\vec{v}=\vec{p}/E^{(0)}_j$, $j=p,e, \mu$ (for leptons (e, $\mu$) $g_s=0$). 
The last equation can be further simplified, noting that the perturbation of the equilibrium distribution function as given in Eq.~(\ref{delphi}) will generate the corresponding perturbation on the fields given by Eq.~(\ref{smalldev}), then:

 \bea
&&\partial_t \delta f^{(j)} + \vec{v} \cdot \nabla_x \delta f^{(j)}
+ \left[ \vec{v}  \times (\nabla_x \times \delta \vec{\cal V}^{(j)} ) + 
  g_s \frac{{M_j^\star}^{(0)}}{{E^{(0)}_j}} \nabla_x \delta \phi \right.\cr
&&
\left. - \partial_t ~\delta \vec{{\cal V}}^{(j)} -\nabla_x \delta {\cal V}_0^{(j)}  
 \right]  \cdot 
 \nabla_p f^{(0)(j)}    
   = 0 \ .
  \label{dvlasov2}
 \eea
Next, we obtain the dispersion relations, starting from the Fourier transform
of the small deviation from equilibrium of the fields and of the distribution functions:
 \begin{equation}
 \left\{
       \begin{array}{c}
          \delta f^{(j)} (\vec{x},\vec{p},t) \\
         \delta \phi (\vec{x},t) \\
         \delta {\cal V}^{(j)}_{\mu} (\vec{x},t)
       \end{array}    
 \right\}  = 
 \int d^3q ~d\omega 
  \left\{
       \begin{array}{c}
          \delta f^{(j)} (\vec{q},\omega,\vec{p}) \\
          \delta \phi (\vec{q},\omega) \\
          \delta {\cal V}^{(j)}_{\mu}(\vec{q},\omega) 
       \end{array}    
 \right\}  e^{i(\omega t - \vec{q} \cdot \vec{x})} \ ,
 \end{equation}
and after substituting the last equation in the Vlasov equation, eq. (\ref{dvlasov2}), one 
obtains for $ \delta f^{(j)}(\vec{q},\omega,\vec{p})$ :
\bwt
\begin{equation}
\delta f^{(j)} = \left[\delta \overrightarrow{\mathcal{V}}^{(j)}-\frac{\left(\delta\mathcal{V}^{(j)}_0-\frac{\vec{p} \cdot \delta \overrightarrow{\mathcal{V}}^{(j)}}{E_{j}^{(0)}}-g_{s} \frac{M_{j}^{\star(0)}}{E_{j}^{(0)}} \delta \phi\right)}{\omega-\frac{\vec{p} \cdot \vec{q}}{E_{j}^{(0)}}} \vec{q}\right] \cdot \nabla_{p} f^{(0)}_{(j)}.
\label{vlasov}
\end{equation}
\ewt
where $\delta f^{(j)}, \mathcal{V}_{\mu}^{(j)}$, and $\delta \phi$ are functions of $(\vec{q}, \omega, \vec{p})$.

We consider the particular case of small perturbations that correspond to longitudinal waves in our reference frame. The longitudinal mode corresponds to small perturbations parallel to $\hat{k}$ and the dispersion relations are obtained by taking $q_{\perp}=0, \; q_{\|} =q$ and $\delta \mathcal{V}_{x}=\delta \mathcal{V}_{y}=0$.
$$
\delta \overrightarrow{\mathcal{V}}^{(j)}=\delta \overrightarrow{\mathcal{V}}_{\|}^{(j)} \equiv \delta \mathcal{V}_{z}^{(j)} \hat{k}, \quad \delta \mathcal{V}_{x}^{(j)}=\delta \mathcal{V}_{y}^{(j)}= \delta \mathcal{V}_{\perp}^{(j)}=0
$$
and,
$$
\delta \overrightarrow{J}^{(j)}=\delta \overrightarrow{J}_{\|}^{(j)} \equiv \delta J_{z}^{(j)} \hat{k}, \quad \delta J_{\perp}^{(j)}=0
$$
From the conservation law, $\partial^{\mu} J_{\mu}^{(j)}=0$, using the Fourier transformation
$\partial^{\mu} J_{\mu}^{(j)}=\partial^{\mu} J_{\mu}^{(j)}(t,\; \vec{x})=0$ for $j=n, p, e, \mu$ leads to the following relation
$$\omega J_0^{(j)}(\vec{q}, \omega)=\vec{q}\cdot\vec{J}^{\;(j)}(\vec{q}, \omega)$$
then,
$$\omega \delta J^{(j)}_0(\vec{q}, \omega)=\vec{q}\cdot\delta \vec{J}^{(j)}(\vec{q}, \omega)=q \delta J_{z}^{(j)}(\vec{q}, \omega).$$

Next, we rewrite the Vlasov equation, Eq.(\ref{vlasov}), for the description of longitudinal mode, and rearrange it as follows:
\bwt
\beq
\delta f^{(j)} =\delta\mathcal{V}^{(j)}_{z}\left(\frac{\partial f^{(0)(j)}}{\partial p_{z}}+\frac{p_{z}}{E^{(0)}_{j}}\frac{\vec{q}\cdot \nabla_{p} f^{(0)(j)}}{\left(\omega-\frac{p_{z} q}{E_{j}^{(0)}}\right)} \right)
-\frac{\delta\mathcal{V}^{(j)}_0\vec{q} \cdot \nabla_{p} f^{(0)(j)}}{\left(\omega-\frac{ p_{z} q}{E_{j}^{(0)}}\right)}
+\frac{M_{j}^{\star(0)} g_{s} \delta \phi \; \vec{q} \cdot \nabla_{p} f^{(0)(j)}}{E_{j}^{(0)}\left(\omega-\frac{p_{z } q}{E_{j}^{(0)}}\right)}, \quad j=p, n.
\label{vlasovL}
\eeq
\ewt

The equations of motion for the mesonic and electromagnetic fields are derived by applying the Euler–Lagrange formalism to the meson Lagrangian. Subsequently, small deviations from the equilibrium field configurations, as defined in Eq.(\ref{smalldev}),  are introduced, and a Fourier transformation is performed, yielding:
\begin{eqnarray}
&& \left[ -\omega^2 + {\vec q}^{~2} + {\tilde m}_s^2 \right] \delta \phi (\vec{q},\omega)
= g_s \sum_{j=p,n} \frac{2 {M^\star}^{(0)(j)}}{(2\pi)^3}\int \frac{d^3 p}{E^{(0)}_j}\delta f^{(j)},  \cr
&&\left[ -\omega^2 + {\vec q}^{~2} + m_v^2 + \frac{\xi}{6}{V^{(0)}_0}^2 +
 2 \Lambda_v {b^{(0)}_0}^2 \right] \delta V_{\mu} +\frac{\xi}{3} {V_0^{(0)}}^2 \delta V_0 \delta_{\mu 0} \cr
&&+ 4 \Lambda_v V_0^{(0)} b^{(0)}_0 \delta b_0 \delta_{\mu 0}
  = g_v  \sum_{j=p,n} \frac{2}{(2\pi)^3}\int \frac{d^3 p}{E^{(0)}_j} p_\mu \delta f^{(j)},\cr
&& \left[ -\omega^2 + {\vec q}^{~2} + m_\rho^2 +
             2 \Lambda_v {V^{(0)}_0}^2 \right] \delta b_{\mu} \cr
&& + 4 \Lambda_v V_0^{(0)} b^{(0)}_0 \delta V_0\delta_{\mu 0}
  = \frac{g_\rho}{2}  \sum_{j=p,n} \tau_j \frac{2}{(2\pi)^3}\int \frac{d^3 p}{E^{(0)}_j} p_\mu \delta f^{(j)},  \cr
&& \left[ -\omega^2 + {\vec q}^{~2} \right] \delta A_{\mu} = \sum_{j=p,e,\mu} Q_j \frac{2}{(2\pi)^3}
  \int \frac{d^3 p}{E^{(0)}_j} p_\mu \delta f^{(j)}, 
  \label{eomf}
\end{eqnarray}
where the effective scalar mass is given by:
\begin{equation}
 {\tilde m}_s^2 = m_s^2 + \kappa \phi^{(0)} + \frac{\lambda}{2}{\phi^{(0)}}^2 - g_s^2
 \sum_{j=p,n} d \rho^{(0)(j)}_s \ ,
\end{equation}
and,
 \begin{equation}
  d\rho_s^{(0)(j)} = \left(\frac{\partial \rho^{(j)}_s}{\partial M^*}\right)_{(0)} = - \frac{2}{(2\pi)^3} \int d^3 p 
  \frac{{\vec{p}}^{~2}} {{E^{(0)}_j}^3}   f^{(0)(j)}~.
 \end{equation}
Next, we use the definition of the current,
\beq
\delta J_{\mu}^{(j)}(\vec{q}, \omega)=\frac{2}{(2 \pi)^{3}} \int \frac{d^{3} p}{E_{j}^{(0)}} p_{\mu} \delta f^{(j)}(\vec{q}, \omega, \vec{p}), \quad j \in(p, n, e, \mu)
\eeq
and write the non-null components of the meson vector field fluctuations as,
\bwt
\bea
&& {\left[-\omega^{2}+\vec{q}^{\;2}+m_{v}^{2}+\frac{\xi}{2} V_{0}^{(0)^{2}}+2 \Lambda_{v} b_{0}^{(0)^{2}}\right] \delta V_{0} +4 \Lambda_{v} V_{0}^{(0)} b_{0}^{(0)} \delta b_{0}=g_{v} \sum_{j=p, n} \delta J_{0}^{(j)}} \cr
&& {\left[-\omega^{2}+\vec{q}^{\;2}+m_{v}^{2}+\frac{\xi}{6} V_{0}^{(0)^{2}}+2 \Lambda_{v} b_{0}^{(0)^{2}}\right] \delta V_{z}=g_{v} \sum_{j=p, n} \delta J_{z}^{(j)}} \cr
&& {\left[-\omega^{2}+\vec{q}^{\;2}+m_{\rho}^{2}+2 \Lambda_{v} V_{0}^{(0)^{2}}\right] \delta b_{0}+4 \Lambda_{v} V_{0}^{(0)} b_{0}^{(0)} \delta V_{0}=\frac{g_{\rho}}{2} \sum_{j=p, n} \tau_{j} \delta J_{0}^{(j)}} \cr
&& {\left[-\omega^{2}+\vec{q}^{\;2}+m_{\rho}^{2}+2 \Lambda_{v} V_{0}^{(0)^{2}}\right] \delta b_{z}=\frac{g_{\rho}}{2} \sum_{j=p, n} \tau_{j} \delta J_{z}^{(j)}},
\eea
\ewt
where, the effective mesons masses are given by,
\bea
\tilde{m}_{\omega}^{2}&=&m_{v}^{2}+\frac{\xi}{2} V_{0}^{(0)^{2}}+2 \Lambda_{v} b_{0}^{(0)^{2}} \cr
\tilde{m}_{\rho}^{2}&=&m_{\rho}^{2}+2 \Lambda_{v} b_{0}^{(0)^{2}}. 
\eea
We use a new definition,
\begin{eqnarray}
\omega^2_s &=&  {\tilde m}_s^2 + {\vec q}^{~2} \nonumber \; ,\\
\omega^2_{\omega} &=& {\tilde m}_{\omega}^2 + {\vec q}^{~2} \nonumber \; ,\\
\omega^2_{\rho} &=& {\tilde m}_{\rho}^2 + {\vec q}^{~2} \nonumber \; .
\end{eqnarray}

\subsection{Dispersion relation}

Following the derivation steps given in Ref.~\cite{rabhi2025}, we obtain the expressions for the density fluctuations of the various species as follows:
\bwt
\bea
 \delta \rho_{j}&=&\frac{1}{2\pi^2}\left\{\sum_{i=p, n}\left[\left(\frac{\omega}{q}\right)^{2} \frac{g_{v}^{2}}{-\omega^{2}+\omega_{\omega}^{2}-\frac{\xi}{3} V_{0}^{(0)^{2}}}-\frac{g_{v}^{2}\left(-\omega^{2}+\omega_{\rho}^{2}\right)}{D(\omega)}\right. \right. 
 +\left(\frac{\omega}{q}\right)^{2} \frac{\left(g_{\rho} / 2\right)^{2} \tau_{i} \tau_{j}}{-\omega^{2}+\omega_{\rho}^{2}}-\frac{\left(g_{\rho} / 2\right)^{2} \tau_{i} \tau_{j}\left(-\omega^{2}+\omega_{\omega}^{2}\right)}{D(\omega)} \cr
&& +\frac{4 \Lambda_{v} V_{0}^{(0)} b_{0}^{(0)} g_{v} \frac{g_{\rho}}{2} \left(\tau_{i}+\tau_{j}\right)}{D(\omega)}  \left.+\frac{\left(g_{s} M_{j}^{\star(0)}\right)^{2}}{\omega_{s}^{2}-\omega^{2}} \frac{1}{E_{F i}^{(0)} E_{F j}^{(0)}}\right] \delta \rho_{i} 
\left.-\frac{1}{\vec{q}^{~2}}\sum_{i=p,e,\mu}Q_i Q_j\delta \rho_{i}\right\}p_{F j} E^{(0)}_{Fj}L\left(s_{j}\right),
\eea
\ewt
where,
$
D(\omega) = (-\omega^2+\omega_\omega^2)(-\omega^2+\omega_\rho^2)
- (4\Lambda_v V_0^{(0)} b_0^{(0)})^2$, $s_j = \frac{\omega}{\omega_{j, F}}$, $\omega_{j, F}=q V_{F_{j}}$, 
$V_{F_j} = \frac{p_{F_j}}{E_{F_j}}$, and $L(s_j)$ is related to the Lindhard function $\Phi$ by
$$
L_j=L(s_j)=2\Phi(s_j)=2-s_j\ln\left(\frac{s_j+1}{s_j-1}\right), \quad \hbox{j=p,n,e},\mu.
$$
\bea
\delta \rho_{i} &=& - \sum_{j=n, p} F^{i,j} L(s_i) \delta\rho_{j} - L(s_i)\sum_{j=e,\mu}C^{i, j}_{A} \delta\rho_{j} , \, i=n,\,p \cr
\cr
\delta\rho_{l} &=& \sum_{k=p,e,\mu}C^{l, k}_{A} L(s_l) \delta\rho_{k}, \; l=e, \mu 
\label{eqq45}
\end{eqnarray}
where,
$$
F^{i,j}=C^{i,j}_{s} - C^{i,j}_{\omega}-C^{i,j}_{\omega\rho}(\tau_i+\tau_j)-C^{i,j}_{\rho}\tau_i\tau_j -C^{i,j}_{A},
$$
with, 
\bea
C^{i,j}_{s} &=&  \frac{1}{2\pi^2}\frac{(g_{s} M_{i}^{\star(0)})^2}{\omega^{2} - \omega^2_s }\frac{p_{F_j}}{E^{(0)}_{F_i}} \cr
C^{i,j}_{\omega} &=& \frac{g^2_{v}}{2\pi^2}\left[\frac{\omega^2-\omega^2_{\rho}}{D(\omega)} -\left(\frac{\omega}{q} \right)^2 \frac{1}{\omega^2-\omega^2_{\omega}+\frac{\xi}{3} {V^{(0)}_0}^2 }\right]p_{F_j} E^{(0)}_{F_j} \cr
C^{i,j}_{\rho} &=& \frac{\left(\frac{g_{\rho}}{2}\right)^2}{2\pi^2}\left[
 \frac{\omega^2-\omega^2_{\omega}}{D(\omega)} 
 -\left(\frac{\omega}{q} \right)^2 \frac{1}{\omega^2-\omega^2_{\rho}}\right]p_{F_j} E^{(0)}_{F_j}\cr
C^{i,j}_{\omega\rho} &=& \frac{1}{2\pi^2}\frac{2 \Lambda_v g_{\omega} g_{\rho} V^{(0)}_0  b^{(0)}_0 p_{F_j} E^{(0)}_{F_j}}{D(\omega)}\cr 
C^{i,j}_{A} &=&-\frac{1}{2\pi^2}\frac{Q_{i}Q_{j}}{{\vec q}^{~2}}p_{F_j} E^{(0)}_{F_j}.
\eea
Equations~\ref{eqq45} expressed in terms of the density fluctuations $\delta\rho_{i}$, can be cast in matrix form as
\bwt
\begin{eqnarray}
\begin{pmatrix}
     1+ F^{pp} L_p    & F^{pn} L_{p} & C^{p,e}_{A} L_{p} &  C^{p,\mu}_{A} L_{p} \\
     F^{np} L_{n}    & 1+ F^{nn} L_n & 0 & 0\\
   C^{e,p}_{A} L_{e}      & 0 & 1- C^{e,e}_{A} L_{e} & -C^{e,\mu}_{A} L_{e} \\
   C^{\mu, p}_{A} L_{\mu}      & 0 & -C^{\mu, e}_{A} L_{\mu} & 1- C^{\mu,\mu}_{A} L_{\mu} \\   
\end{pmatrix} 
    \begin{pmatrix}
    \delta \rho_{p} \\
    \delta \rho_{n} \\
    \delta\rho_{e} \\
 \delta\rho_{\mu} \\ 
\end{pmatrix} =    \mathbf{M}(\omega,q)
    \begin{pmatrix}
    \delta \rho_{p} \\
    \delta \rho_{n} \\
    \delta\rho_{e} \\
 \delta\rho_{\mu} \\
 \end{pmatrix}  =0.   
\label{tab:my_label}
\nonumber
\end{eqnarray}
\ewt
with,
\begin{equation}
M_{ij} = \delta_{ij} + L(s_i)\,\mathcal{C}_{ij}(\omega,q),
\end{equation}
The dispersion relation is given by:
\bwt
\begin{equation}
(1 - L_e C_A^{e,e})(1 - L_\mu C_A^{\mu,\mu})
\Big[
(1 + L_p F^{pp}_{\text{eff}})(1 + L_n F^{nn})
- L_p L_n F^{pn} F^{np}
\Big] = 0
\label{eq:det_factorized}
\end{equation}
\ewt
Only the proton-proton component of the hadronic subspace is modified 
by the leptonic couplings, resulting in the effective replacement
\bwt
\begin{equation}
F^{pp} \rightarrow F^{pp}_{\text{eff}} = 
F^{pp}
- \frac{
(1 - L_\mu C_A^{\mu,\mu}) L_e C_A^{p,e} C_A^{e,p}
+ (1 - L_e C_A^{e,e}) L_\mu C_A^{p,\mu} C_A^{\mu,p}
+ L_\mu L_e C_A^{p,e} C_A^{\mu,e} C_A^{\mu,p}
}{
(1 - L_e C_A^{e,e})(1 - L_\mu C_A^{\mu,\mu})
}.
\label{eq:Fpp_eff}
\end{equation}
\ewt

From the final equation, the dispersion relation is obtained by evaluating the determinant of a four-dimensional linear system involving density fluctuations $\delta \rho_{p}$, $\delta \rho_{n}$,  $\delta \rho_{e}$, and  $\delta \rho_{\mu}$. The eigenmodes $\omega$ of the system correspond to the values for which the determinant vanishes - that is, the roots of the matrix determinant.

\section{Results and discussion}
The present analysis builds upon our earlier work~\cite{rabhi2025}, in which the collective modes of asymmetric nuclear matter were explored.
In the present work,  we present and analyze the collective longitudinal modes of neutron star matter in $\beta$-equilibrium, that may propagate in the outer core of the star. Two compositions are considered: nucleonic matter consisting of neutrons, protons, and electrons (npe matter), and matter that additionally includes muons (npe$\mu$ matter). Both compositions are relevant for the outer core of neutron stars, where the densities are sufficiently high for muon formation via weak processes to become energetically favorable. The inclusion of muons modifies the chemical equilibrium conditions and the composition of the medium, thereby affecting the behavior of collective excitations. Muons typically appear at densities below nuclear saturation density.

For simplicity, hyperonic degrees of freedom are not included in the present study, although they are expected to become relevant at the higher baryon densities considered. The role of hyperons will be explored in a future work.

We employ three representative relativistic mean-field (RMF) models in our analysis, whose saturation properties are listed in Table~\ref{tab:nuclear}: NL3~\cite{nl3}, NL3$\omega\rho$~\cite{Pais16, Horowitz01}, and FSU2H~\cite{tolos17, tolos17_2, Negreiros18}. The parameterizations NL3 and NL3$\omega\rho$ are characterized by a large incompressibility and  a relatively stiff equation of state (EOS) at high densities, due to the absence or the small coupling of the quartic term $\omega^4$. However, NL3$\omega\rho$ has a much softer symmetry energy compared to NL3.  In contrast, the FSU2H model includes  additional non-linear couplings that soften the EOS. Its symmetry energy is also quite soft compared with NL3. These models span a representative range of symmetry energy behaviors, its slope, and the density dependence of pressure and energy density, providing a suitable framework for a systematic study of their impact on collective modes in dense matter.
\begin{table}[htb]
 \begin{tabular}{lccccccc}
\hline\hline
  & $\rho_0$ &$M^*/M$ & $B$ & $K$ & $Q_0$ &$E_\mathit{sym}$   &$L$  \\
& $ [\mathrm{fm}^{-3}]$& & [MeV]& [MeV]& [MeV]&[MeV]& [MeV]  \\\hline 
NL3 &0.1480& 0.596 &-16.24&271 & 197.9&37.4&118.0   \\
NL3$\omega\rho$&0.1480& 0.596 &-16.24&271& 197.9&31.5&55.0\\
FSU2H&0.1505& 0.593 &-16.28&238&-246.7 &30.5 &44.5 \\
\hline
\end{tabular}
\caption{Nuclear matter properties of the RMF models considered in this study.}
\label{tab:nuclear}
\end{table}

\begin{figure}[t]
\includegraphics[width=1.0\linewidth,angle=0]{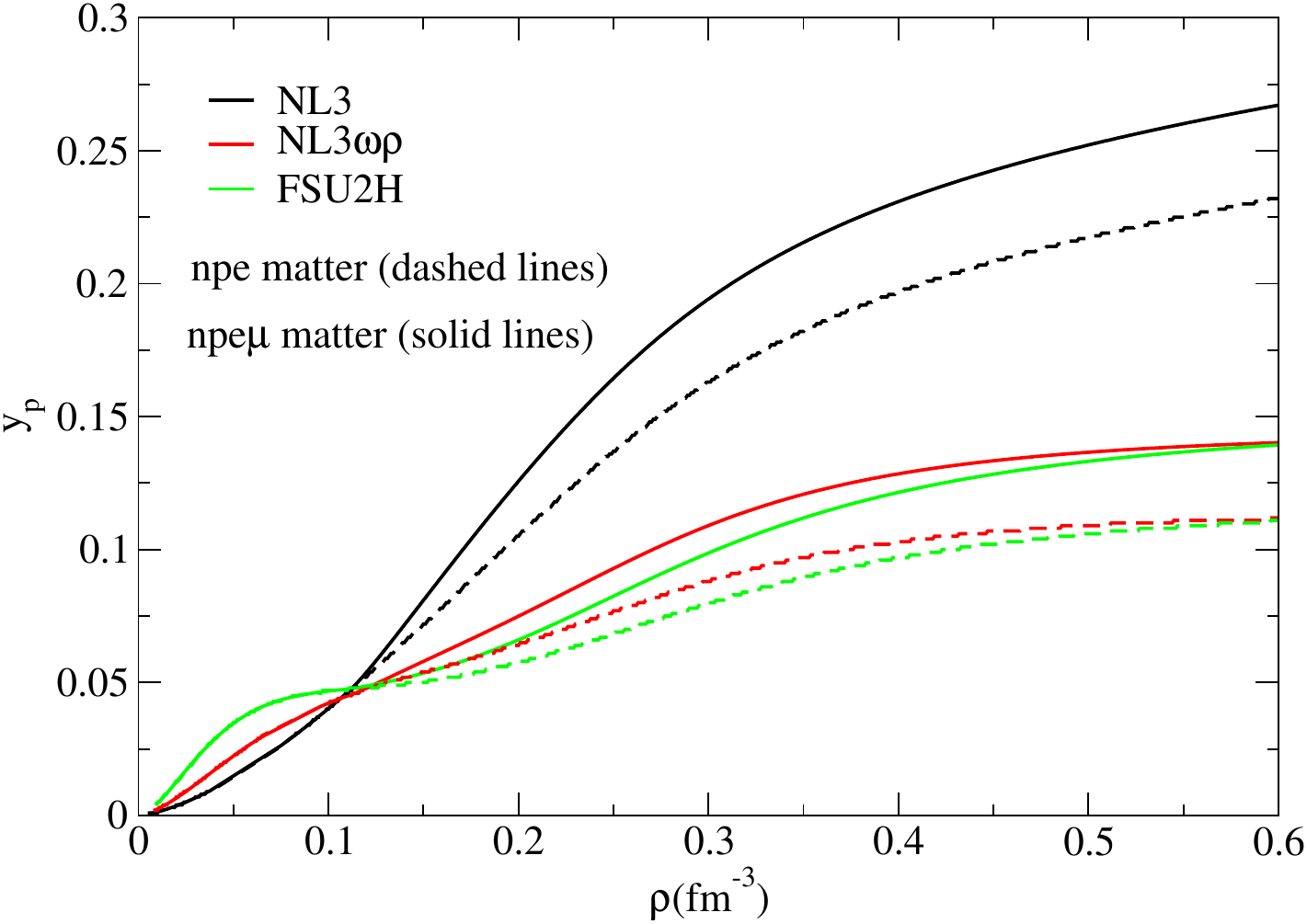}
\caption{Proton fractions as function of baryon density, for neutral $\beta$-equilibrated neutron star matter. Dashed curves correspond to npe matter, while solid curves represent npe$\mu$ matter.}
\label{figure1}
\end{figure}

In Fig.~\ref{figure1}, the proton fraction of $\beta$-equilibrated neutron star matter is plotted as a function of baryon density for the three parametrizations considered. Dashed lines correspond to npe matter, whereas solid linees represent npe$\mu$ matter. The inclusion of muons in charge-neutral, $\beta$--equilibrated matter increases the proton fraction at high densities by modifying the chemical equilibrium conditions. The proton fraction reflects directly the density behavior of the symmetry energy: a large  symmetry energy as the one of NL3 disfavors very asymmetric neutron star matter.  Due to its stiff symmetry energy above saturation density, NL3 model predicts the largest proton fractions over most of the density range, reaching values above $y_p\simeq 0.25$ for npe$\mu$ matter and $y_p\simeq 0.225$ for npe matter at around three times saturation density in charge-neutral, $\beta$-equilibrated matter. In contrast, the proton fraction of  both NL3$\omega\rho$ and FSU2H saturates at approximately 
$y_p\simeq 0.14$ for npe$\mu$ matter and $y_p\simeq 0.12$ for npe matter. At subsaturation densities, the trend reverses: FSU2H with the smallest symmetry energy slope at saturation predicts the largest proton fraction, while NL3, that has the largest slope $L$, gives the smallest proton fractions. This behavior is consistent with the properties of the three EOSs.  

\begin{figure*}[ht]
\includegraphics[width=1.0\linewidth,angle=0]{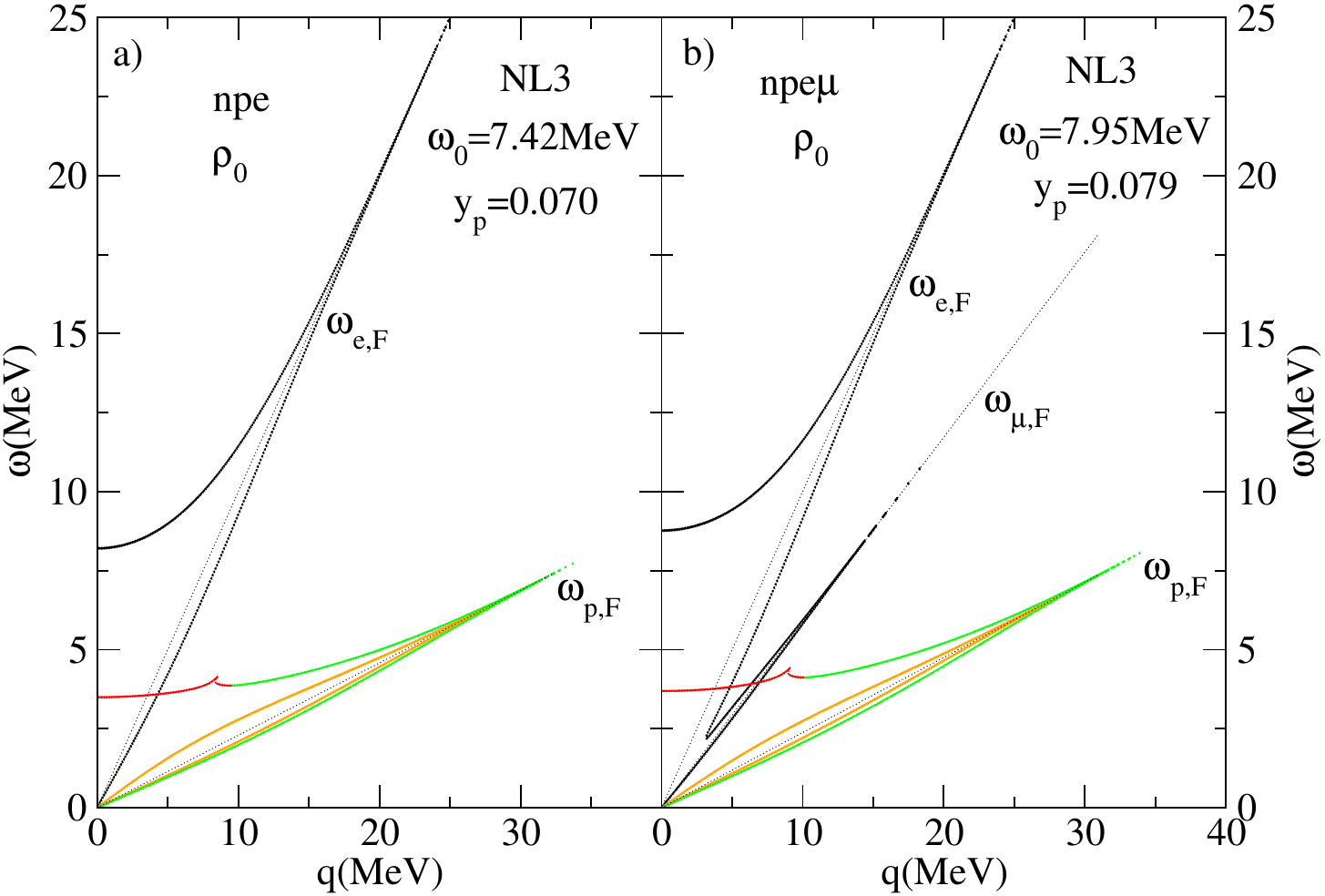}
\caption{
Collective modes as a functions of the momentum transfer $q$ at baryon density $\rho=\rho_0$, calculated within NL3 model. Results are shown for $\beta$-equilibrated npe matter (left panel) and npe$\mu$ matter (right panel). Black lines correspond to isovector modes, and orange lines to isoscalar modes in $\beta$-equilibrated npe and npe$\mu$ matter. Red lines denote isovector modes, and green lines denote isoscalar modes for np matter including Coulomb effects. The Fermi modes $\omega_{i, F} = qV_{F_i}, (i=p, n, e, \mu)$ are indicated by black dotted lines for reference. The plasmon mode-frequency at zero-momentum transfer, $\omega_{0}$, for a relativistic degenerate electron (and muon) gas is provided. The proton fraction at the chosen baryon density is also shown for reference.} 
\label{fig2a}
\end{figure*}

The detailed characterization of the longitudinal collective excitation spectrum of electrically neutral, $\beta$-equilibrated nuclear matter is determined within a covariant Vlasov framework based on the relativistic mean-field models introduced. The analysis enables us to disentangle the respective roles of $\beta$ equilibrium, Coulomb interactions, and the proton fraction in shaping the dispersion relations, and to assess their sensitivity to the density dependence of the symmetry energy encoded in the NL3, NL3$\omega\rho$, and FSU2H parametrizations. Particular emphasis is placed on comparisons between npe (and npe$\mu$) matter in $\beta$ equilibrium and np matter with Coulomb interactions included at a fixed proton fraction, allowing us to disentangle purely nuclear effects from electromagnetic contributions. 
The emergence of a plasmon-like mode together with a nuclear collective mode of predominantly proton-like character highlights the nontrivial interplay between charged degrees of freedom and nuclear dynamics in dense matter. As discussed in \cite{Baldo:2008pb}, the coupling of the electrons to the protons transforms the proton mode  a plasmon-like mode (in a uniformly charged background)  into a sound-like mode. This is clearly seen in the Fig. \ref{fig2a}: the red curve corresponding to the plasmon-like proton mode in pn matter shifts to the orange sound-like mode when dynamical electron are included in the dispersion relation.

\begin{figure*}[ht]
\begin{tabular}{cc}
\includegraphics[width=0.5\linewidth,angle=0]{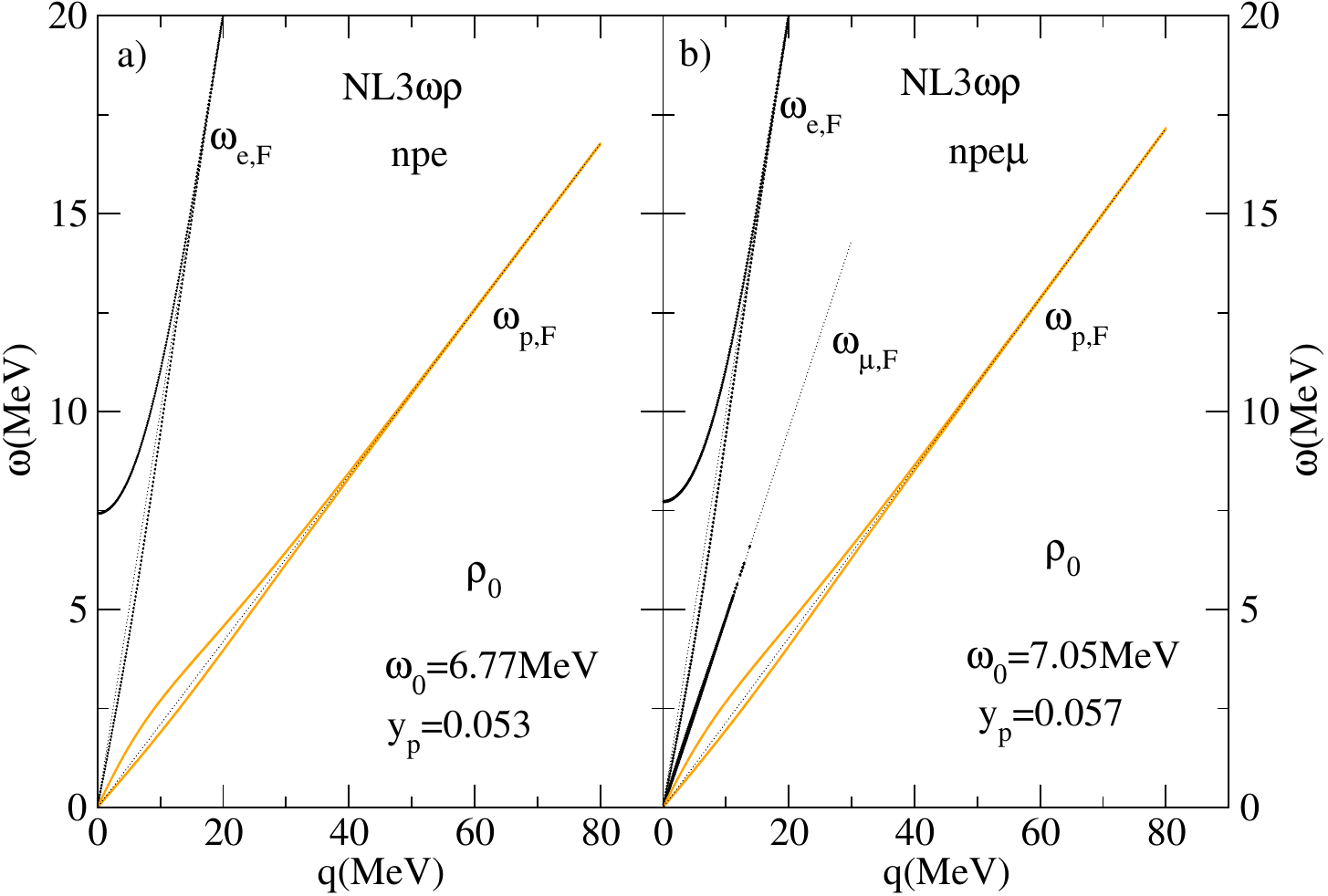} &
\includegraphics[width=0.5\linewidth,angle=0]{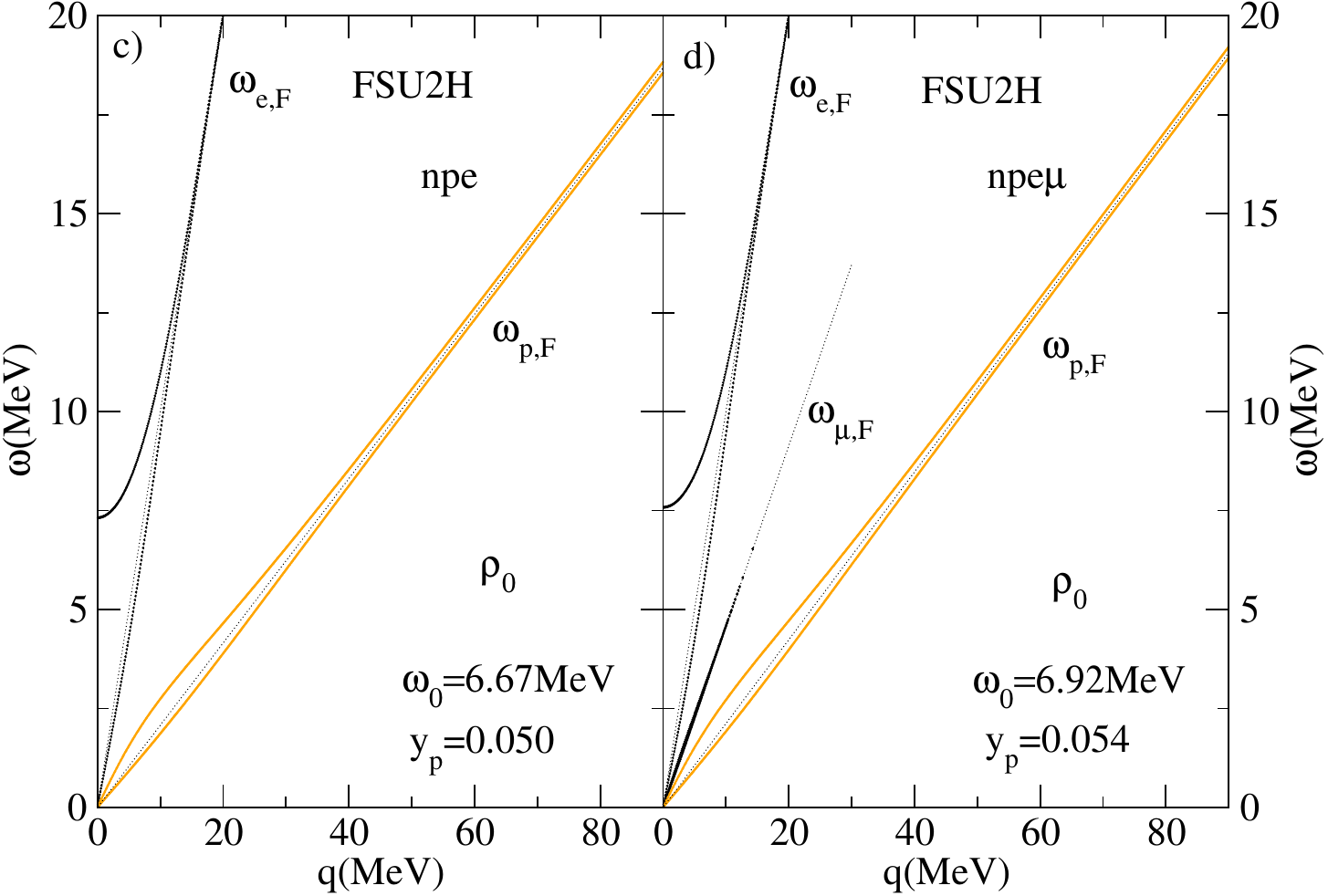} 
\end{tabular}
\caption{
Collective modes as a functions of the momentum transfer $q$ at baryon density $\rho=\rho_0$, calculated within NL3$\omega\rho$ (left panels ) and FSU2H (right panels) models. Results are shown for $\beta$-equilibrated npe matter (left panel) and npe$\mu$ matter (right panel). Black lines correspond to isovector modes, and orange lines to isoscalar modes in $\beta$-equilibrated npe and npe$\mu$ matter. The Fermi modes $\omega_{i, F} = qV_{F_i}, (i=p, n, e, \mu)$ are indicated by black dotted lines for reference. The plasmon mode frequency at zero-momentum transfer, $\omega_{0}$, for a relativistic degenerate electron (and muon) gas is also included. The proton fraction at the chosen baryon density is provided for reference.} 
\label{fig2b}
\end{figure*}

We first examine the dependence of the collective-mode energies on the momentum transfer in electrically neutral matter, $\beta$-equilibrated neutron star matter. Figures~\ref{fig2a} and~\ref{fig2b}  show this dependence  at saturation density, $\rho=\rho_0$. The plasmon frequency at zero-momentum transfer, $\omega_{0}$, is indicated in each figure. For a relativistic degenerate electron gas, $\omega_{0}=\left(e^2\rho_e/E_{F_{e}}\right)^{1/2}$, while in the presence of muons it becomes $\omega_{0}=\left(e^{2}\left(\rho_{e}/E_{F_{e}}+\rho_{\mu}/E_{F_{\mu}}\right)\right)^{1/2}.$ The proton fraction, obtained from the same $\beta$-equilibrated calculation of electrically neutral matter at the same density, is given as a label  for reference. Results for npe (left panels) and npe$\mu$ (right panels) matter are shown. The solid black lines correspond to the plasmon modes (electronic in npe matter and leptonic in npe$\mu$ matter) and to the nuclear isovector modes, while the solid orange lines represent the nuclear isoscalar modes.  In addition, results for np matter in a uniformly negative charged background   including Coulomb effects are displayed in both panels, with solid red lines denoting isovector modes and solid green lines denoting isoscalar modes.  As noted in Ref.~\cite{avancini05}, collective modes are classified through the ratio of proton to neutron density fluctuations, $\delta\rho_p/\delta\rho_n$: positive values correspond to isoscalar-like modes (in-phase oscillations), while negative values indicate isovector-like modes (out-of-phase oscillations). 

At saturation density, the collective response of $\beta$-equilibrated npe matter is characterized by the presence of two well-separated modes for all three models. The higher-energy branch corresponds to a plasmon-like mode dominated by electron oscillations, originating from the long-range Coulomb interaction, while the lower-energy branches correspond to nuclear collective modes of predominantly proton-like character. The plasmon-like mode ceases to propagate beyond a certain $q_{\text{max}}$, while the nuclear modes persist only up to model-dependent limits: about $30~\mathrm{MeV}$ for NL3, $80~\mathrm{MeV}$ for NL3$\omega\rho$, and above $90~\mathrm{MeV}$ for FSU2H. At this density the neutron-like mode does not propagate. Comparisons with $np$ matter including Coulomb effects show that electromagnetic interactions control the low-momentum behavior and generate the isovector plasmon-like branch for the protons.

The nuclear response splits into Landau-damped and undamped modes. In $\beta$-equilibrated $npe$ matter, both are isoscalar for all models. In contrast, for $np$ matter with Coulomb interactions, the undamped mode changes character: it is mainly isovector at low momentum and becomes isoscalar above $q \approx 10~\mathrm{MeV}$, whereas the damped mode remains isoscalar across the full momentum range. Including muons introduces an additional plasmon-like branch. In $npe\mu$ matter, two distinct leptonic plasmons appear, associated with electron and muon oscillations, alongside a proton-dominated nuclear mode. This reflects the presence of two charged leptonic species with different masses that couple weakly via the Coulomb field, while baryonic dynamics remain largely unaffected.

A comparison of the NL3, NL3$\omega\rho$, and FSU2H models at saturation density $\rho_0$ reveals a common qualitative structure of collective excitations in npe and npe$\mu$ matter, compare Fig. \ref{fig2b} with Fig. \ref{fig2a}. All three parameterizations predict plasmon-like leptonic modes and a pair of proton-like nuclear modes of zero-sound character, consisting of Landau-damped and undamped branches. When only Coulomb interactions are included, the Landau-undamped proton mode acquires an isovector character and exhibits a plasmon-like behavior at low momentum, while evolving toward an isoscalar nature at larger $q$. In contrast, the Landau-damped mode remains isoscalar and retains its zero-sound character throughout the entire momentum range. Quantitatively, NL3 yields the softest response, with proton-like modes limited to $q_{max}\simeq 30-35$MeV and maximum energies $\omega_{max}\simeq 7-8$MeV, whereas NL3$\omega\rho$ and FSU2H predict significantly stiffer nuclear dynamics, with proton-like modes extending to $q_{max} \simeq 80 - 90$MeV and reaching $\omega_{max} \geqslant 17-19~\text{MeV}$. Between these two, FSU2H generally produces slightly higher excitation energies, while NL3$\omega\rho$ exhibits comparable momentum extensions. Both models predict similar proton fractions about one fourth smaller than the one of NL3. This explains the differences found on the modes frequency between NL3 and the other two. The inclusion of muons does not qualitatively modify the nuclear collective modes or their isospin character; instead, it introduces an additional low-energy plasmon-like branch associated with muon excitations, while leaving the electron plasmon and the relative model trends essentially unchanged. Minor quantitative shifts induced by the modified $\beta$-equilibrium proton fraction have only a weak impact on the Coulomb-coupled proton mode at low momenta. 

\begin{figure*}[t]
\includegraphics[width=1.0\linewidth,angle=0]{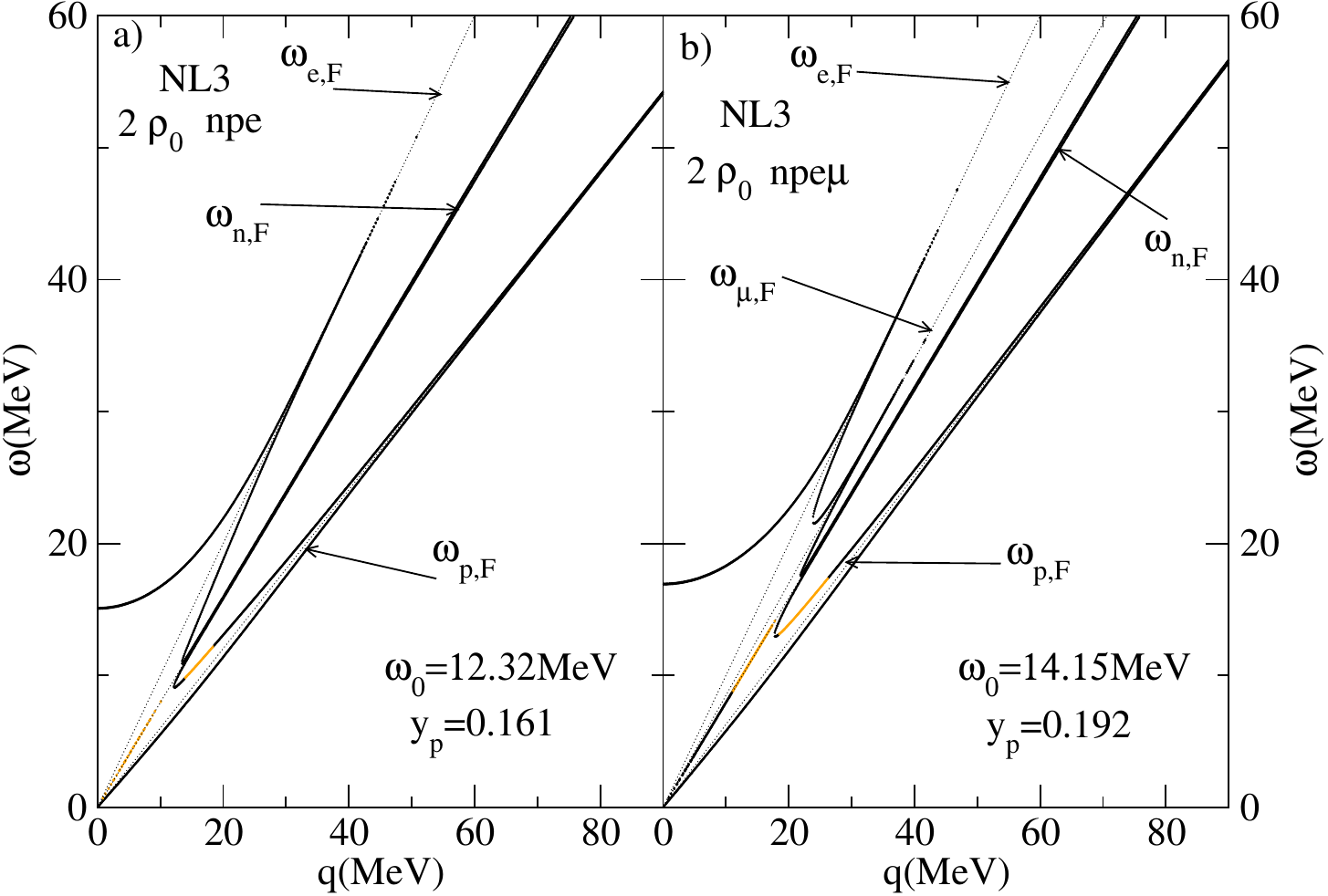} 
\caption{
The same as Fig.~\ref{fig2a}, but for baryon density of $\rho=2\rho_0$ and without results for np matter including Coulomb effects.} 
\label{fig2d}
\end{figure*}

\begin{figure*}[t]
\begin{tabular}{cc}
\includegraphics[width=0.5\linewidth,angle=0]{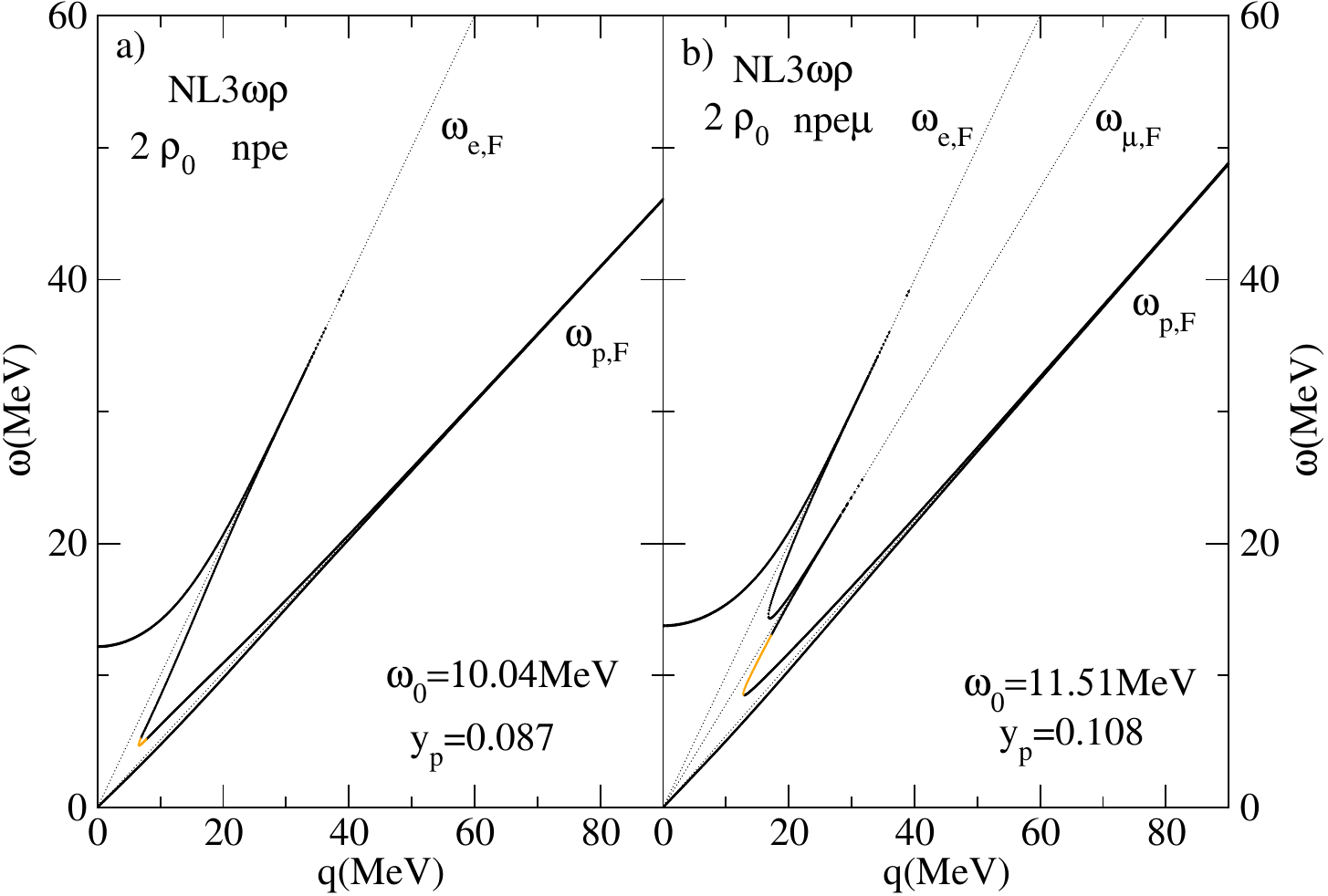} & 
\includegraphics[width=0.5\linewidth,angle=0]{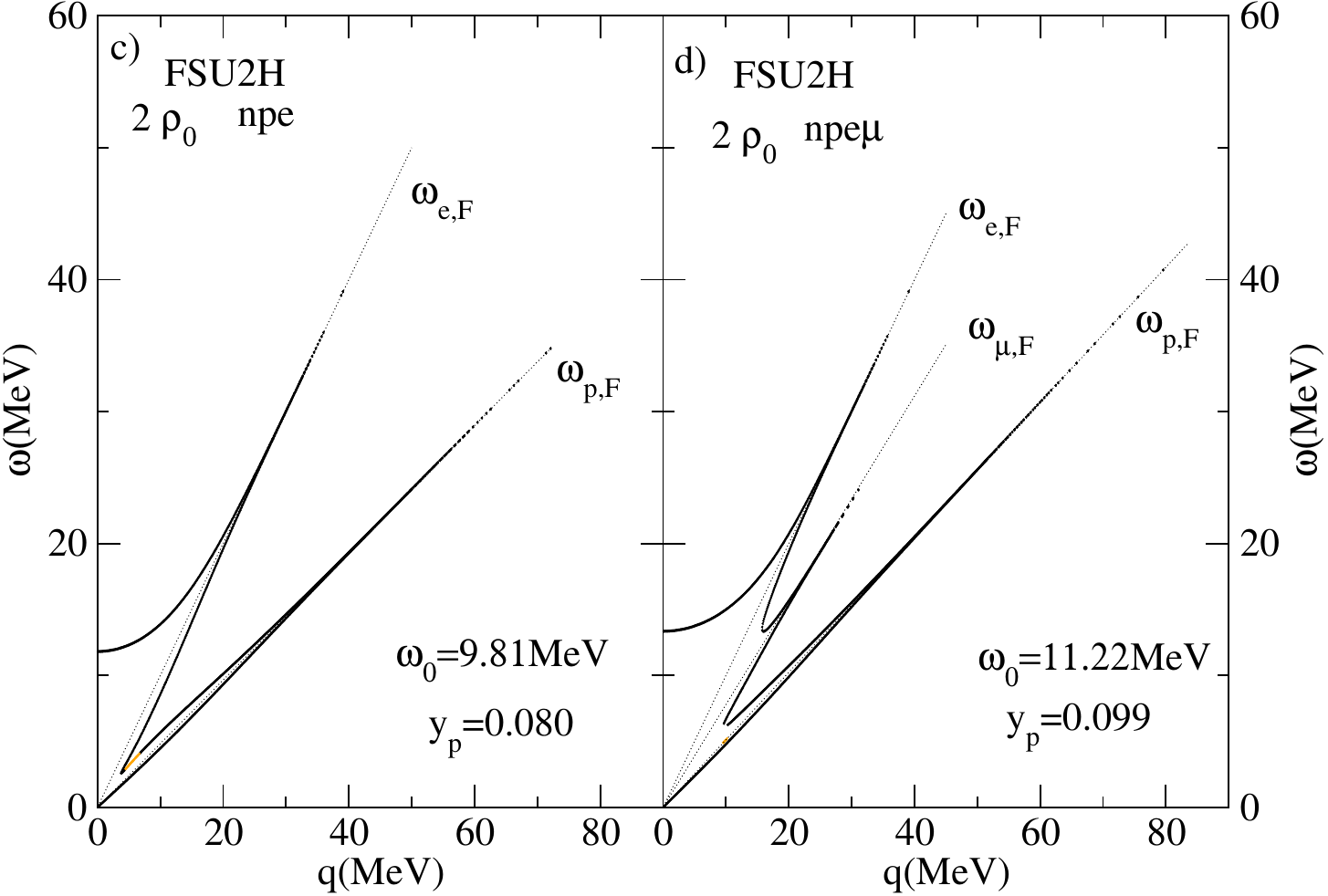} 
\end{tabular}
\caption{
The same as Fig.~\ref{fig2b}, but for baryon density of $\rho=2\rho_0$.}
\label{fig2e}
\end{figure*}

We next examine the dependence of the collective–mode energies on the momentum transfer  at twice the saturation density, $\rho=2\rho_0$. 
Figures~\ref{fig2d} and~\ref{fig2e} display the corresponding results for NL3 and NL3$\omega\rho$, FSU2H. At twice saturation density, the collective modes display an increased sensitivity to the nuclear interaction, in particular, to the density dependence of the symmetry energy and stiffness of the EOS. For all models, a plasmon-like mode, involving the electron or the electron and muon, and a proton-dominated nuclear mode persist. In addition, a distinct zero-sound neutron-dominated collective mode emerges in the NL3 model. 

This neutron-like branch corresponds to a mode that propagates in isoscalar stiff models as discussed in \cite{greco2003,avancini05}. In \cite{rabhi2025},  it was shown that soft EOS did not present this mode.  However, if stiff enough the mode propagates and acquires an isoscalar character above $\sim 2 \rho_0$.

For the NL3 model, we recall that in our previous study the following behavior was identified: for np matter neglecting the proton charge, two pairs of sound-like modes appear above and below the lines ($\omega_{F_{i}}$, i=n,p); when the Coulomb interaction is included, one of the proton modes becomes a plasmon-like mode while the other turns into a strongly damped sound mode; finally, in npe matter, a plasmon-like mode above the electron plasmon is found, together with two pairs of sound-like modes located above and below the lines ($\omega_{F_{i}}$, i=n,p). We noted that some of these modes do not propagate at low momenta. 

\begin{figure*}[t]
\includegraphics[width=1.0\linewidth,angle=0]{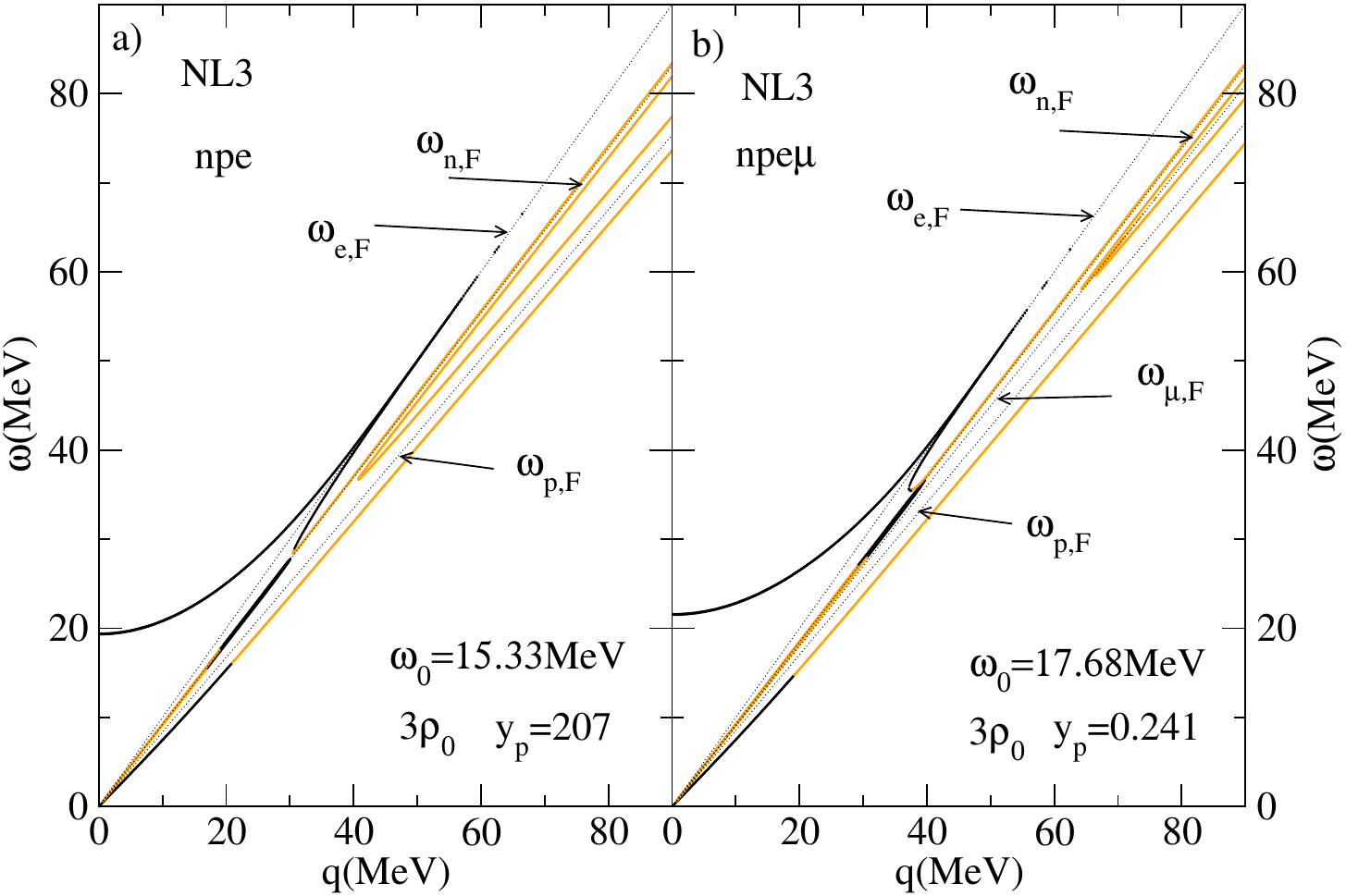} 
\caption{
The same as Fig.~\ref{fig2a}, but for baryon density of $\rho=3\rho_0$. Results for np matter including the Coulomb interaction are omitted for clarity.} 
\label{fig2f}
\end{figure*}

\begin{figure*}[t]
\begin{tabular}{cc}
\includegraphics[width=0.5\linewidth,angle=0]{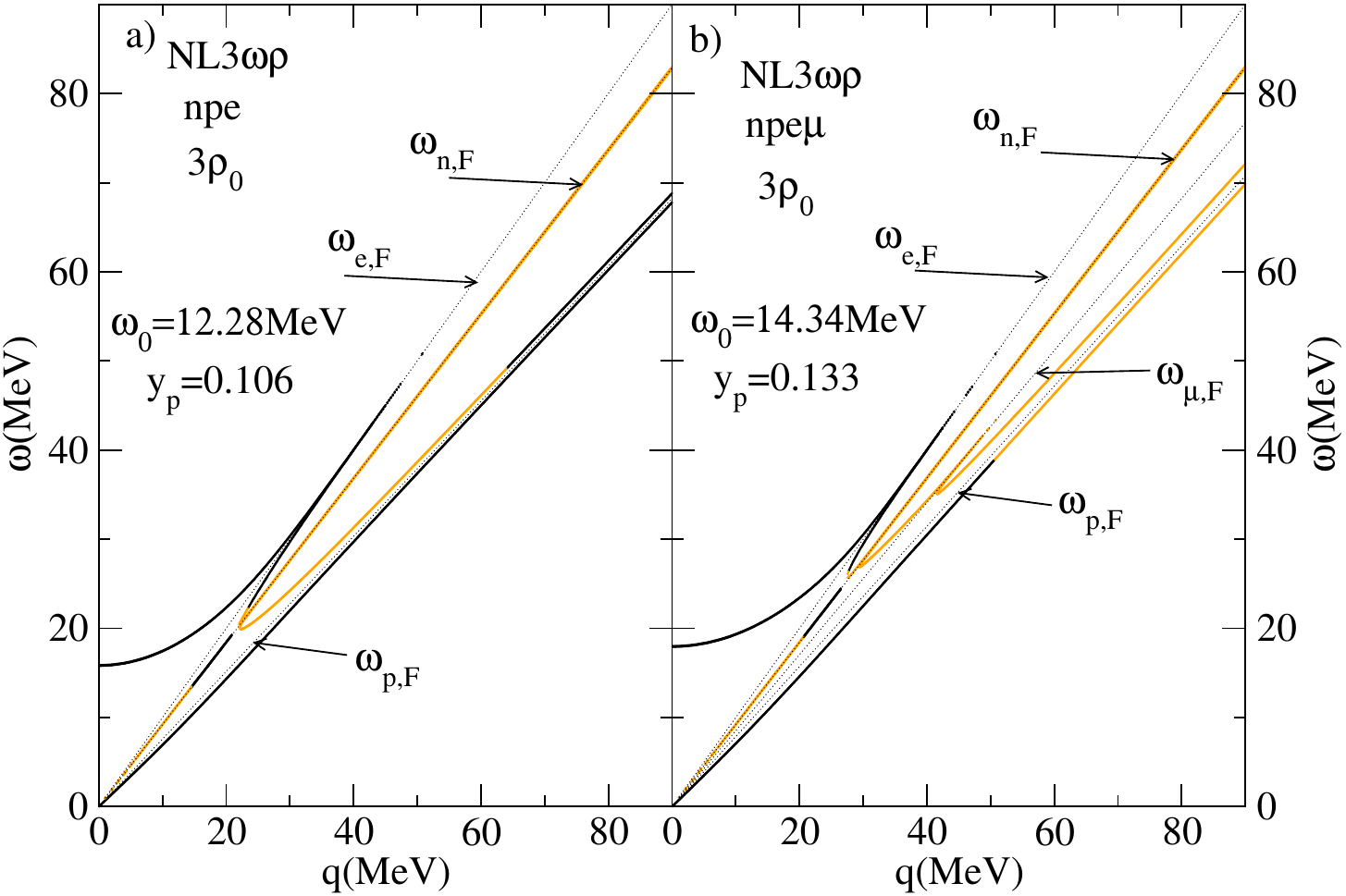} & 
\includegraphics[width=0.5\linewidth,angle=0]{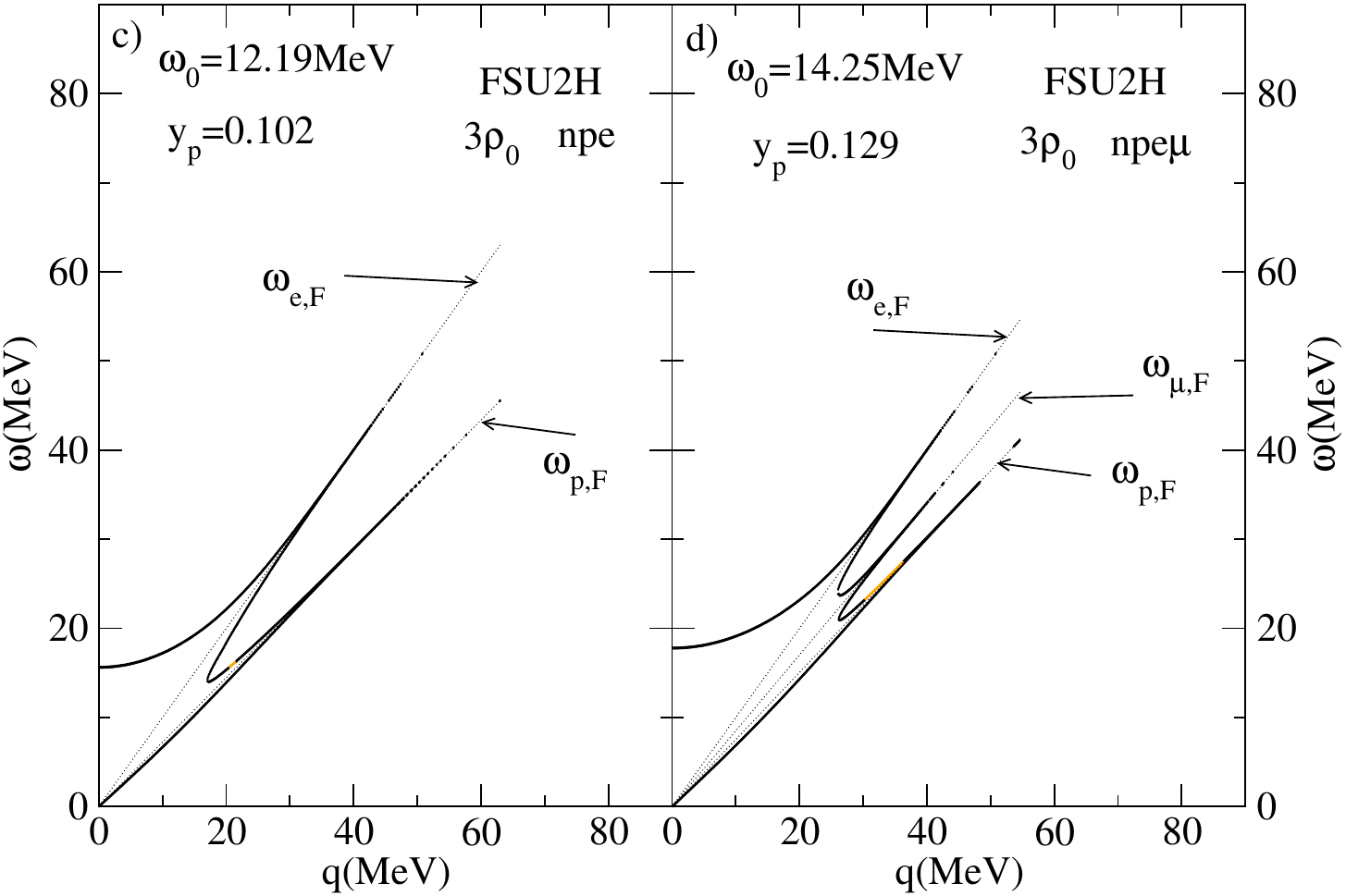} 
\end{tabular}
\caption{
(Color online) The same as Fig.~\ref{fig2b}, but for baryon density of $\rho=3\rho_0$. Results for np matter including the Coulomb interaction are omitted for clarity.}
\label{fig2g}
\end{figure*}

We now consider the case of three times the saturation density, $\rho=3\rho_0$. Figures~\ref{fig2f} and~\ref{fig2g} depict the corresponding results. The collective response becomes more sensitive to the properties of the different  models. For the NL3 model, the spectrum of $\beta$-equilibrated npe matter consists of a plasmon-like mode together with multiple nuclear branches. Both neutron-like and proton-like modes are present and appear as pairs of Landau-damped and undamped solutions, reflecting the pronounced stiffness of the interaction and the resulting decoupling of neutron and proton oscillations at high density.
The inclusion of muons gives rise to an additional plasmon-like mode of muon character, while leaving the nuclear sector essentially unchanged. At low momenta, the neutron-like mode is isoscalar up to $q\simeq 20$ MeV, where it becomes isovector and propagates up to $q\simeq 32$ MeV; beyond this point, it reverts to an isoscalar character and continues to propagate.

For the NL3$\omega\rho$ model, the collective spectrum at $3\rho_0$ is less intricate. In npe matter, one plasmon-like mode, a neutron-like mode, and a pair of proton-like modes exhibiting Landau-damped and undamped behavior are observed. As at lower densities, the presence of muons leads to the appearance of an extra plasmon-like branch without qualitatively modifying the nuclear modes. 

In contrast, the FSU2H model yields a markedly simpler collective response at $3\rho_0$. In $\beta$-equilibrated npe matter, only one plasmon-like mode is found, which is strongly coupled to the proton-like nuclear mode. This behavior reflects the soft high-density symmetry energy of FSU2H and the resulting strong coupling between leptonic and baryonic degrees of freedom. When muons are included, the collective response exhibits the corresponding plasmon-like mode associated with the leptonic sector, with no additional distinct nuclear branches developing.

Overall, the present results reveal a clear evolution of the collective excitation spectrum with increasing density. At saturation density, the response is dominated by leptonic plasmons and proton-like nuclear modes, with muons introducing additional plasmon branches. At intermediate densities, the possible emergence of neutron-dominated modes is highly model dependent and directly linked to the stiffness of the symmetry energy. At supra-saturation densities, stiff interactions such as NL3 give rise to a rich spectrum of neutron- and proton-like modes, including Landau-damped branches, whereas softer models such as FSU2H exhibit a reduced and more strongly coupled collective response.

The strong sensitivity of the collective modes to composition, density, and the isovector sector of the nuclear interaction has important implications for neutron-star physics. In particular, these modes influence the dynamical response of dense matter and may significantly affect neutrino propagation, scattering, and damping in neutron-star interiors. The appearance of neutron-dominated modes and the degree of Landau damping at supra-saturation densities can modify neutrino opacities and transport properties, thereby impacting the thermal evolution and cooling of neutron stars. Collective excitations in $\beta$-equilibrated matter thus provide a valuable probe of the high-density behavior of the symmetry energy and its role in neutron-star phenomenology.

\begin{figure*}[t]
\begin{tabular}{cc}
\includegraphics[width=0.5\linewidth,angle=0]{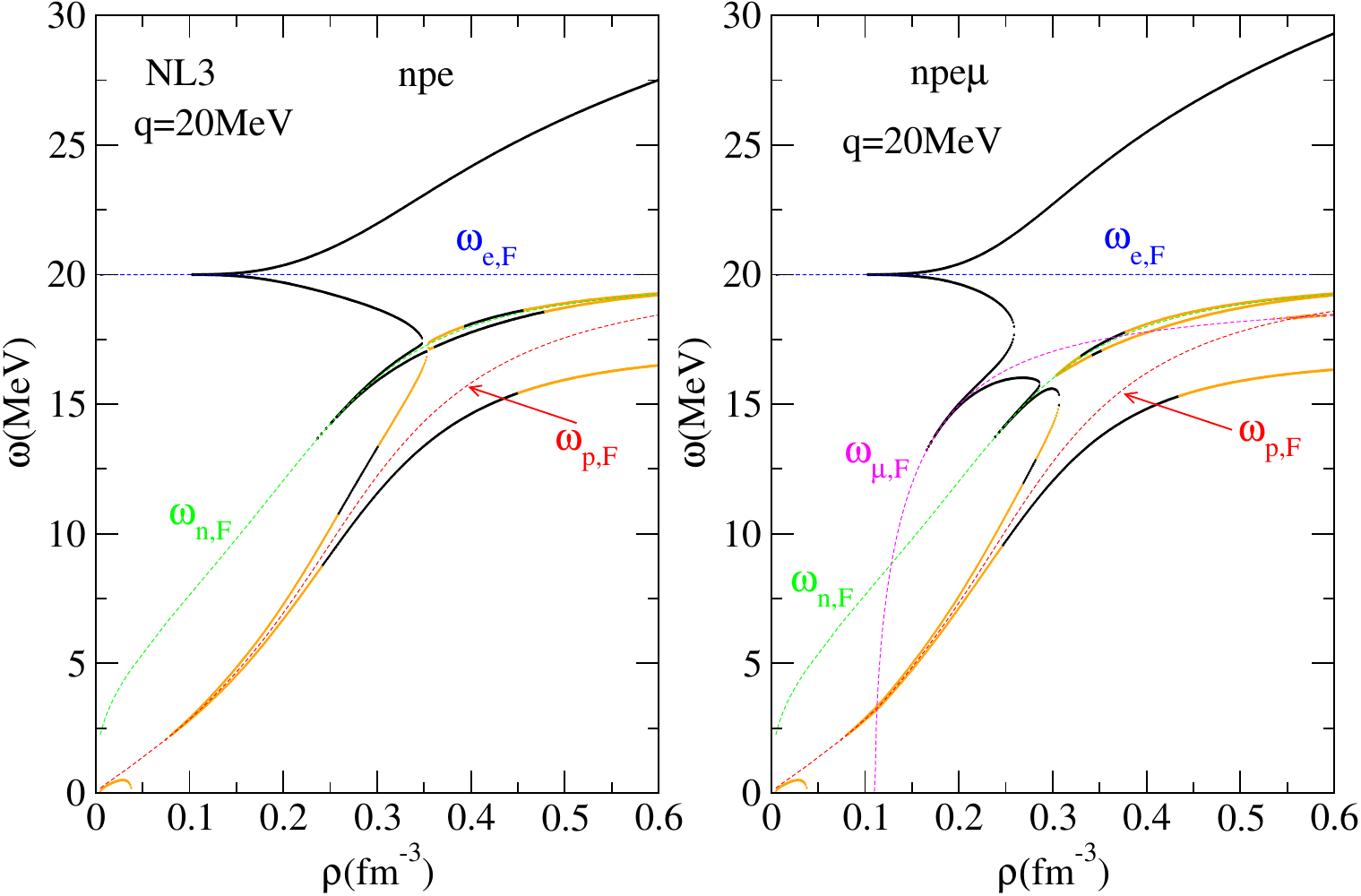}&\includegraphics[width=0.5\linewidth,angle=0]{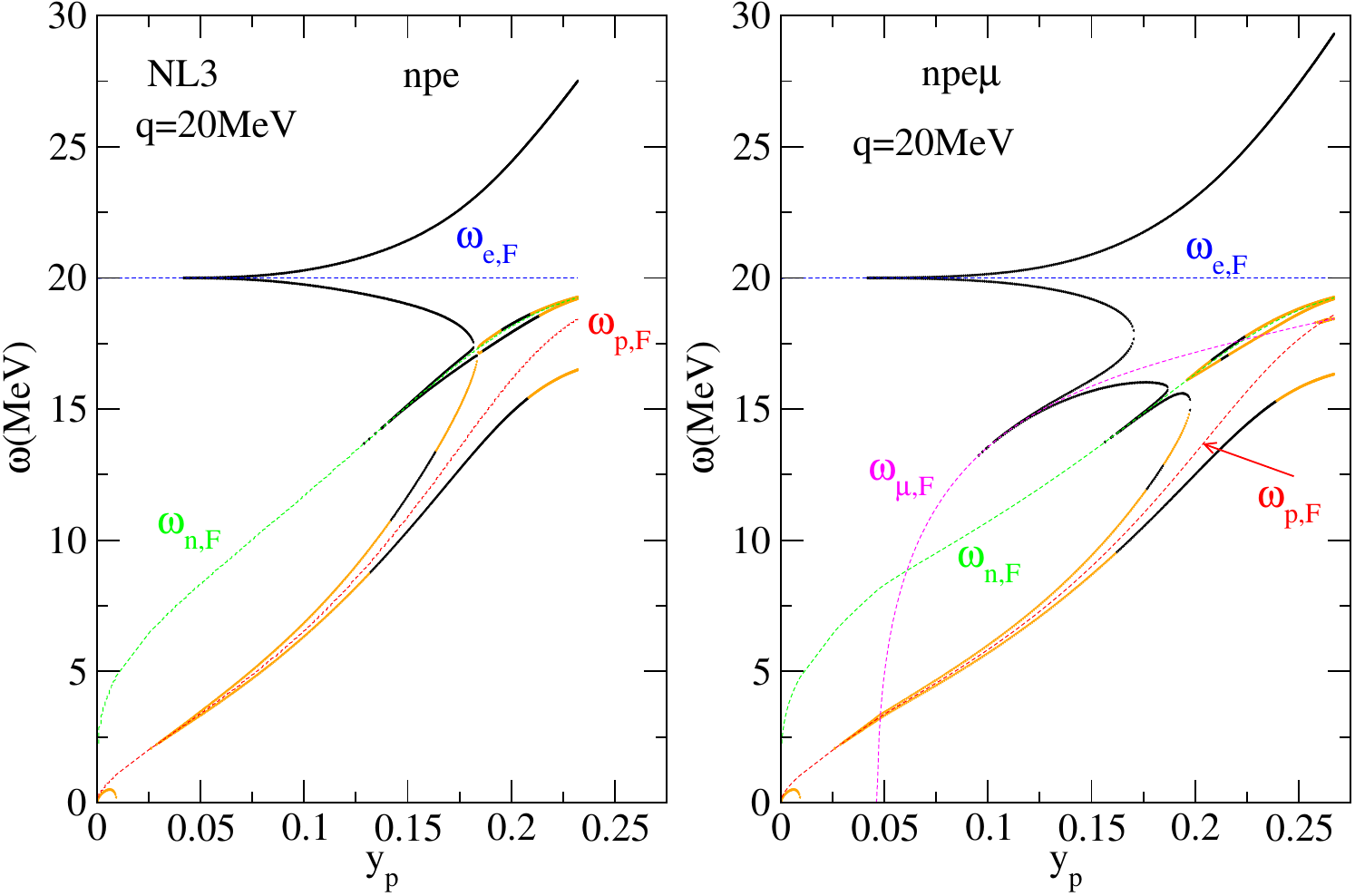}\\
\includegraphics[width=0.5\linewidth,angle=0]{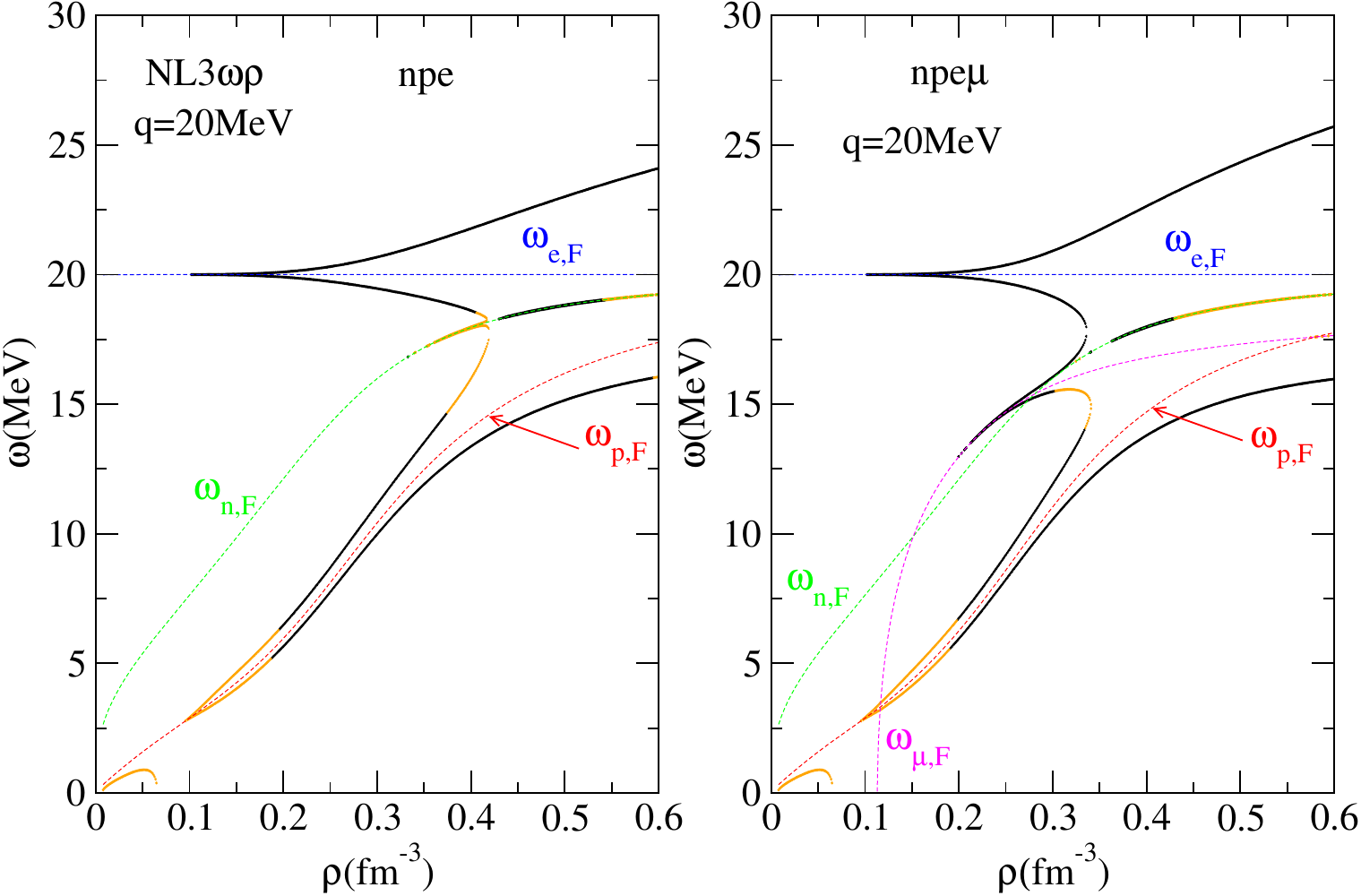}&\includegraphics[width=0.5\linewidth,angle=0]{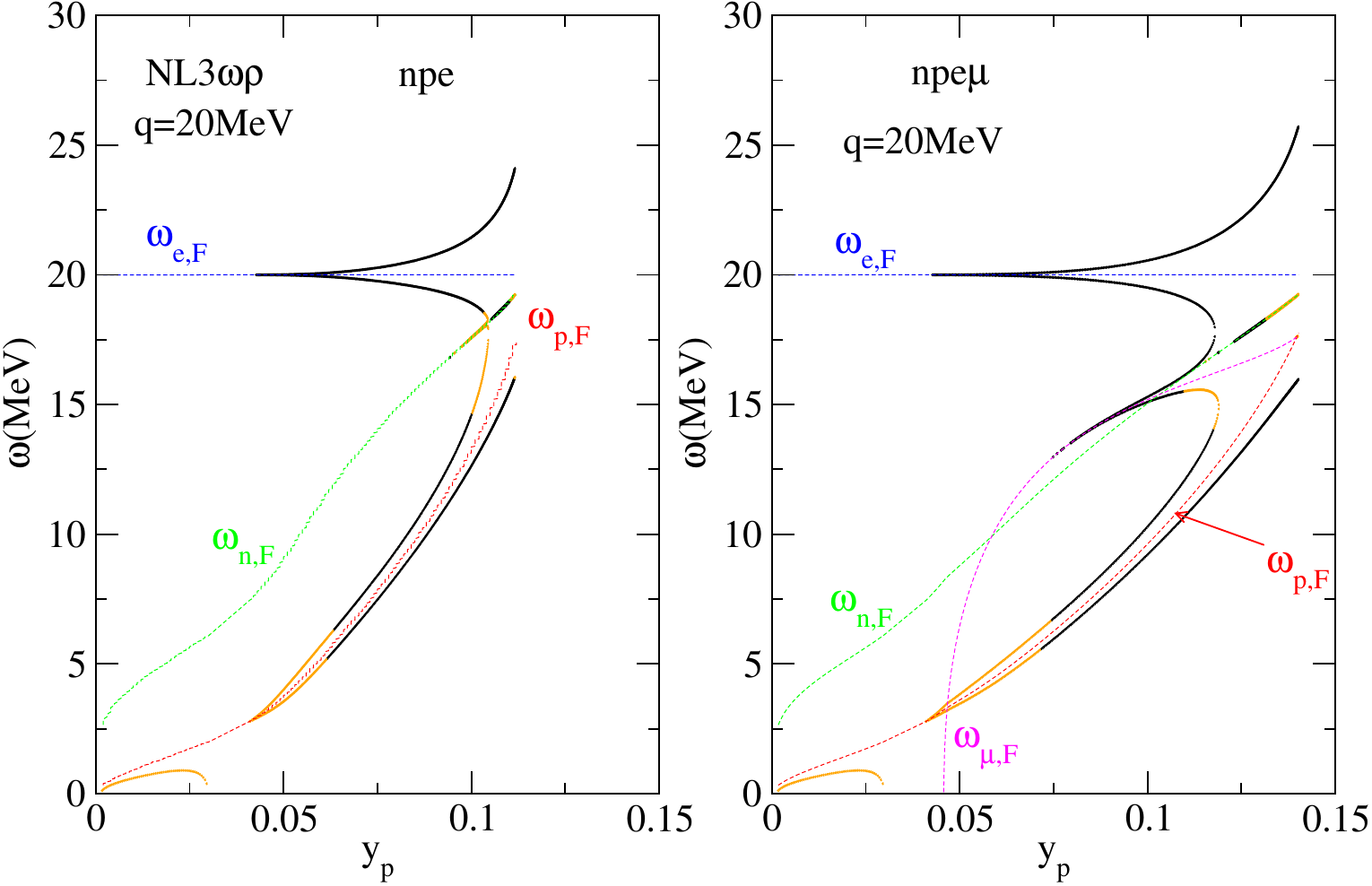}\\
\includegraphics[width=0.5\linewidth,angle=0]{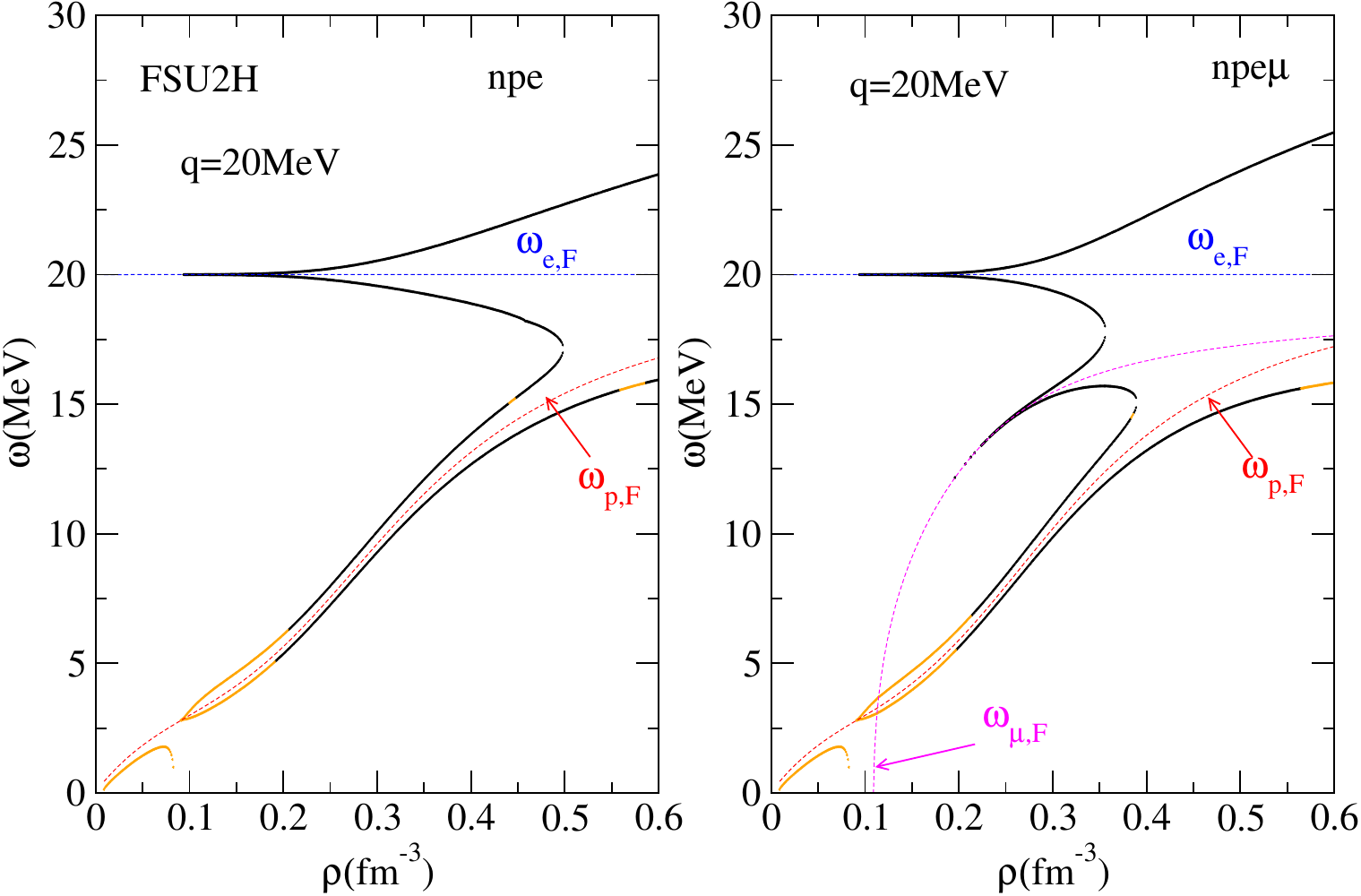} &\includegraphics[width=0.5\linewidth,angle=0]{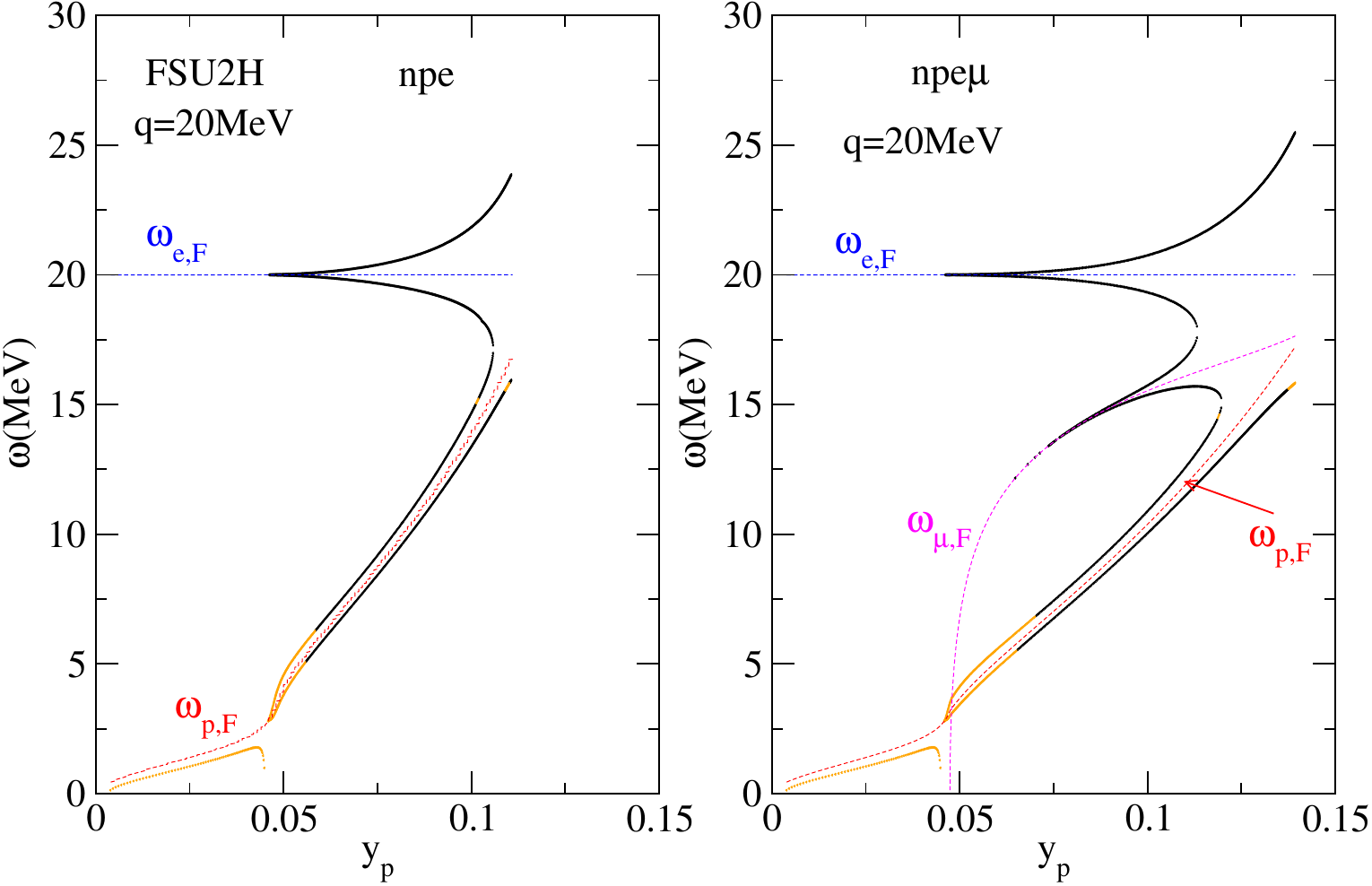}
\end{tabular}
\caption{
Collective–mode energies $\omega$ (MeV) at $q=20$ MeV. Left (right) panels show the dependence on baryon density (proton fraction). Results are given for NL3 (top), NL3$\omega\rho$ (middle), and FSU2H (bottom) models, for charge-neutral $\beta$-equilibrated npe (left subpanels) and npe$\mu$ (right subpanels) matter. Isovector (isoscalar) modes are shown by black (orange) solid lines. The corresponding Fermi modes, $\omega_{i,F}=qV_{F_i}$, are plotted as dotted lines for protons (red), neutrons (green), electrons (blue), and muons (magenta).} 
\label{fig3a}
\end{figure*}

We now investigate the dependence of the collective–mode energies on both the baryon density and the proton fraction resulting from the $\beta$-equilibrium condition, at fixed momentum transfer. In Figure~\ref{fig3a}, the left panels show the density dependence of the collective excitation energies $\omega$ (in MeV), while the right panels display the corresponding dependence on the proton fraction, for a momentum transfer equal to $20$~MeV. The analysis is performed within the NL3 (top row), NL3$\omega\rho$ (middle row), and FSU2H (bottom row) relativistic mean-field models. 
For each model, the results are shown for charge-neutral $\beta$-equilibrated npe matter (columns 1 and 3) and for npe$\mu$ matter (columns 2 and 4). The collective modes are classified according to their isospin character: isovector modes are represented by black solid lines, whereas isoscalar modes are shown by orange solid lines. For comparison, the corresponding single-particle Fermi modes, $\omega_{i,F}=qV_{F_i}$, are also included and plotted as dotted lines. Specifically, red, green, blue, and magenta dotted lines denote the proton, neutron, electron, and muon Fermi modes, respectively. Some conclusions that can be drawn: i) at low densities all figures show a mode that comes alone and identifies the crust  region, below 0.1 fm$^{-3}$ for $\beta$-equilibrium matter; ii) if no muons are included the proton mode couples to the electron mode above 0.3-0.4fm$^{-3}$  and does not propagate at higher densities; iii) including muons, these particles couple to the electron mode  and the proton mode and do not propagate above these densities, 0.25-0.3 fm$^{-3}$; iv)  as discussed before FSU2H presents no neutron-like collective mode that propagates, due to the softness of the EoS; v) NL3 is presenting the most complex set of modes and also is the model that reaches larger proton fractions due to its stiff isospin behavior. Both NL3$\omega\rho$ and FSU2H reach similar proton fractions, their different behavior being due to the stiffer isoscalar behavior of NL3$\omega\rho$.

\begin{figure*}[t]
\begin{tabular}{cc}
\includegraphics[width=0.5\linewidth,angle=0]{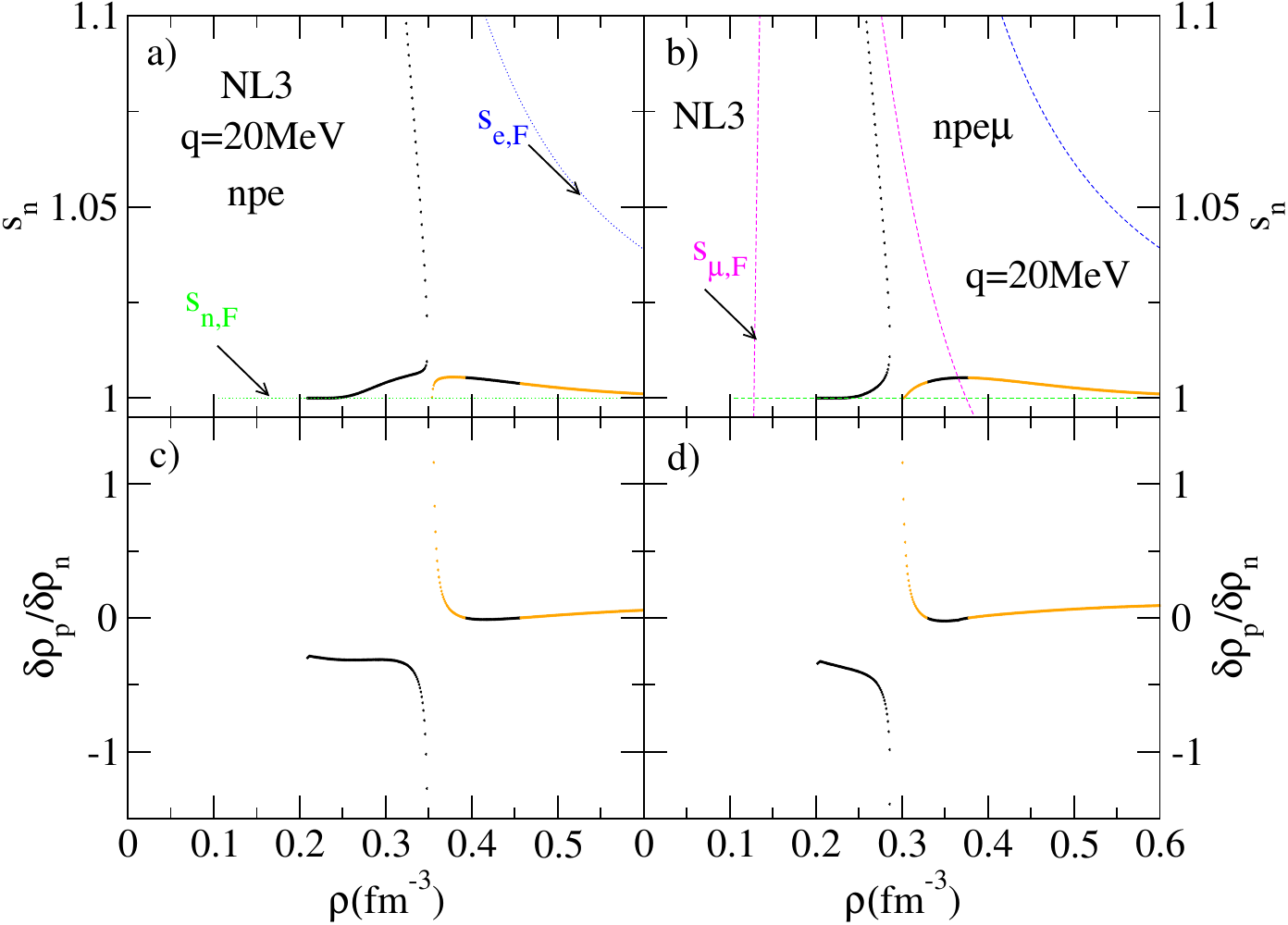} &
\includegraphics[width=0.5\linewidth,angle=0]{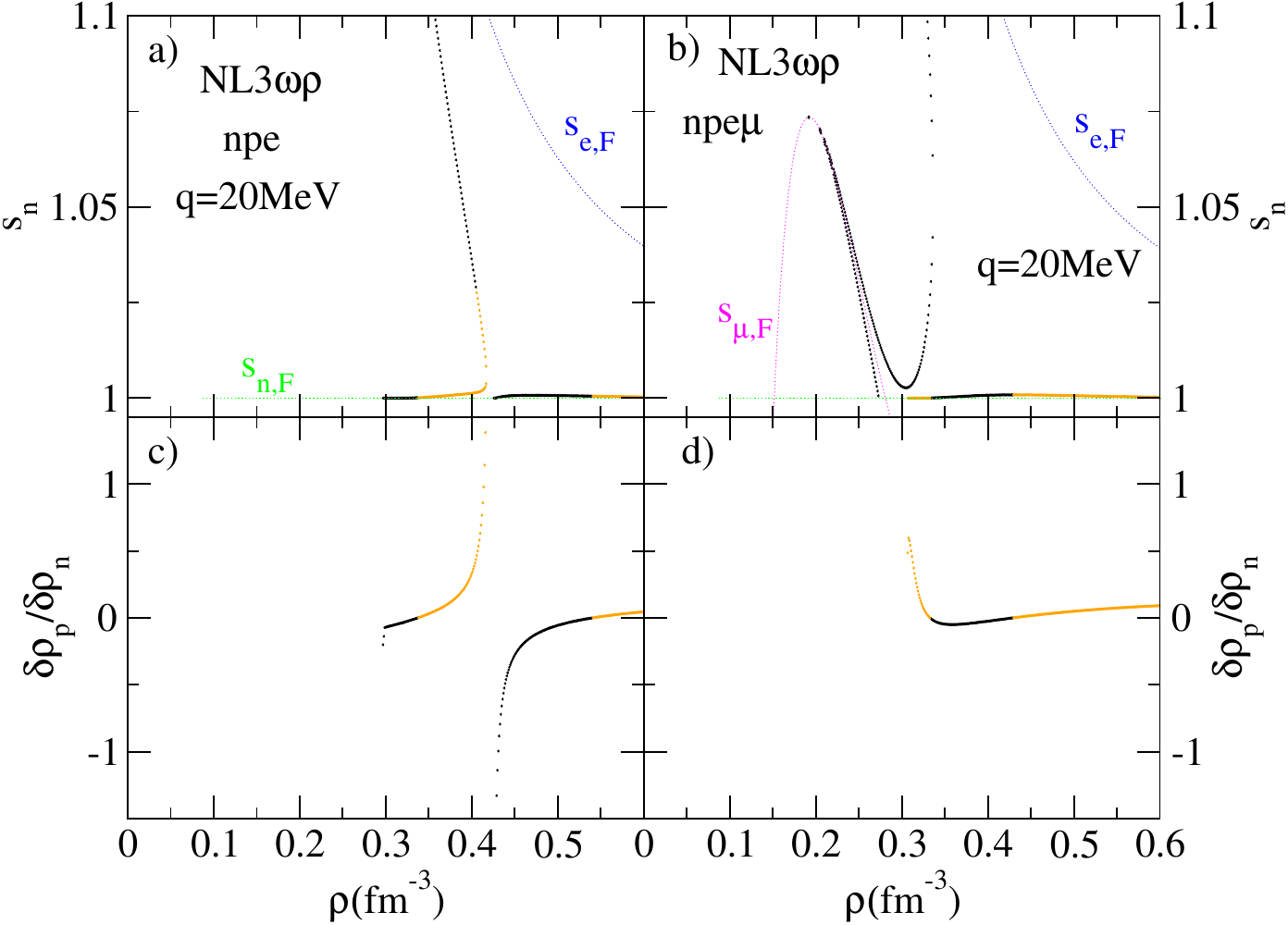} \\
\includegraphics[width=0.5\linewidth,angle=0]{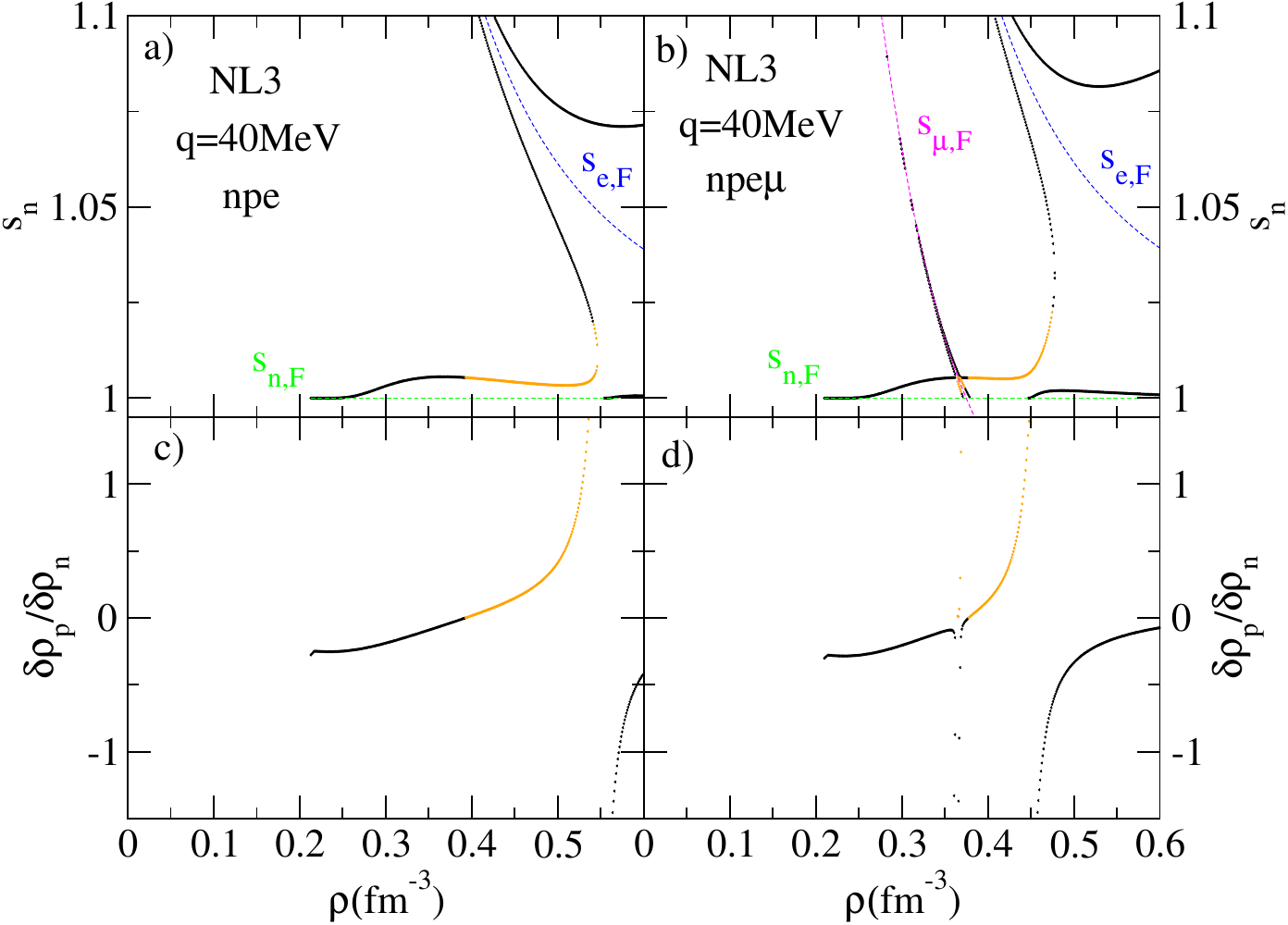} &
\includegraphics[width=0.5\linewidth,angle=0]{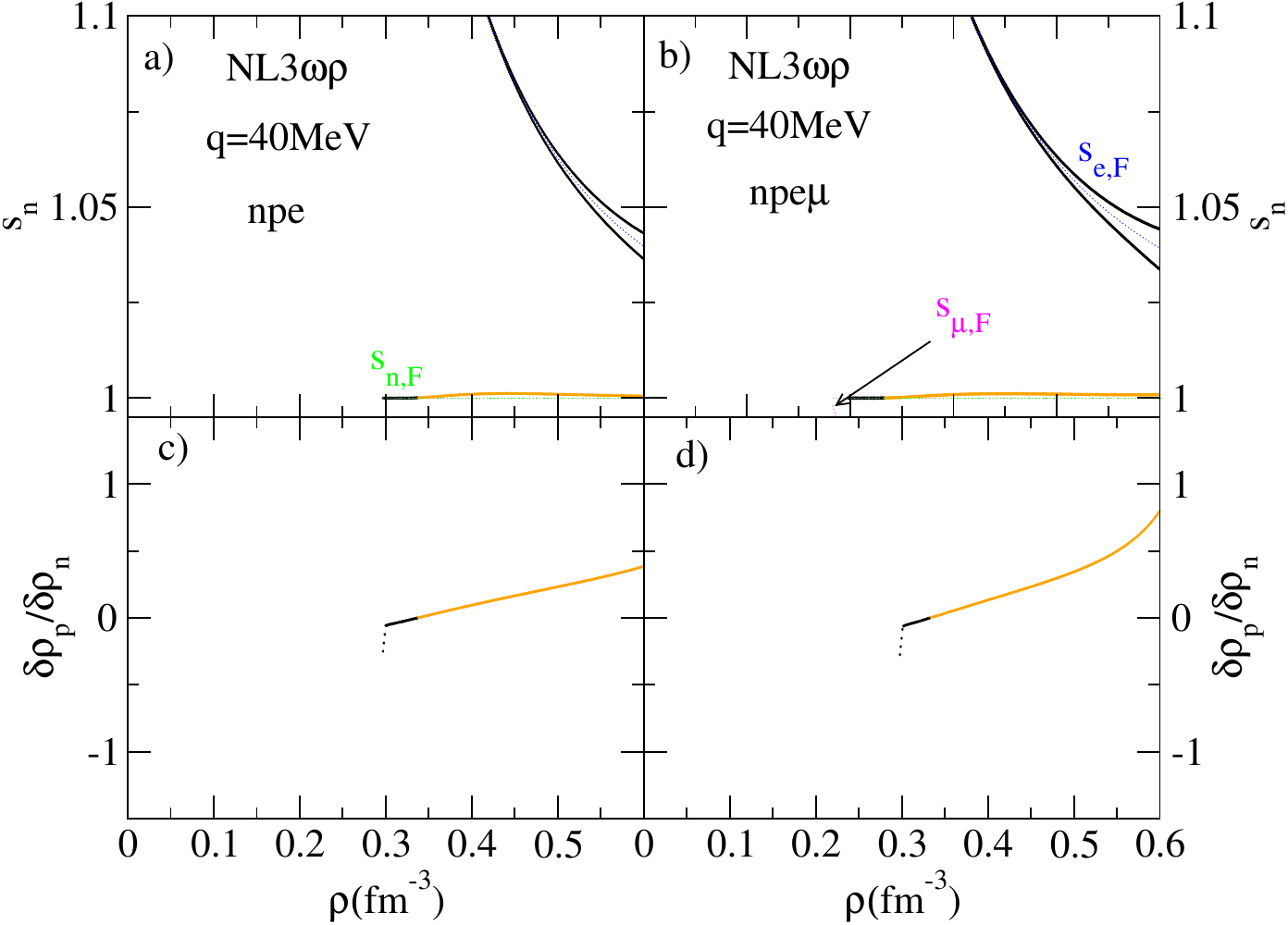}
\end{tabular}
\caption{Sound velocity of the collective modes,
$s_n=\omega/\omega_{n, F}$, together with the ratios of density
fluctuations $\delta\rho_p/\delta\rho_n$ as function of the baryon
density. Results are shown for the NL3 model in the left panel for 
momentum transfers $q=20$~MeV (top quartet), and $q=40$~MeV (bottom quartet), and for the NL3$\omega\rho$ model in the right panel  for $q=20$~MeV (top quartet), and $q=40$ MeV (bottom quartet).
For each quartet, the left panels correspond to npe matter in
$\beta$ equilibrium, while the right panels correspond to
npe$\mu$ matter. For reference, the Fermi sound velocities of the
individual particle species are also shown.
}
\label{fig4a}
\end{figure*}

Having analyzed the longitudinal collective–mode energies as functions of baryon density and momentum transfer, we now focus on the neutron sector and consider the corresponding sound velocity, defined by $s_n=\omega/\omega_{n, F}$, where $\omega_{n, F}=q V_{Fn}$ denotes the neutron Fermi frequency. This quantity characterizes the propagation of neutron-dominated modes with respect to the neutron Fermi surface and provides a useful criterion to identify their dynamical regime. In particular, values $s_n>1$ correspond to modes located outside the neutron particle–hole continuum, while $s_n<1$ indicates modes subject to Landau damping. 
In Fig.~\ref{fig4a},  the sound velocity $s_n$ and the corresponding ratios $\delta\rho_p/\delta\rho_n$ are  shown as functions of the baryon density for two values of the momentum transfer. Results are presented for the NL3 (left panel) and NL3$\omega\rho$ (right panel) models, for $q=20$  and $40$ MeV (from top to bottom), and for both npe and npe$\mu$ matter. For reference, the Fermi sound velocities of the individual particle species are also displayed. The FSU2H parametrization is omitted, as no neutron-like collective modes are found within this model.

For $q=20$ MeV there is still a strong coupling between the plasmon modes and the nucleon modes, while for $q=40$ MeV this coupling occurs above 3 to four times saturation density for NL3 and well above these densities for NL3$\omega\rho$. In NL3, the coupling to the plasmon mode  is strong and in the presence of the muon it moves to lower densities and energies as could already be seen in Fig. \ref{fig3a}. For NL3$\omega\rho$ the coupling occurs for npe matter. For npe$\mu$ matter, the coupling of the plasmon mode occurs directly with the proton like mode. The neutron like mode does not propagate at densities below two times saturation density.  The character of the nucleon mode at the coupling is of the isovector like type but switches to isoscalar like just above the coupling. 
The stiffer symmetry energy of NL3 leads to larger proton fractions in $\beta$-equilibrium, enhancing lepton densities and promoting earlier and stronger coupling to plasmon modes. Conversely, NL3$\omega\rho$ produces smaller proton fractions, which reduces the lepton contribution and shifts or suppresses the coupling.

\section{Conclusions and outlooks}
In this work, extending previous studies based on the covariant formulation of the Vlasov equation, we investigated the collective modes of neutron-star matter composed of neutrons, protons, electrons, and muons within a relativistic mean-field (RMF) framework. Both nuclear and leptonic excitations were analyzed in charge-neutral, $\beta$-equilibrated matter for a representative set of RMF models commonly employed in neutron-star studies. These models differ in key properties, including the stiffness of the equation of state of symmetric nuclear matter and the density dependence of the symmetry energy. Our results show that the propagation of collective excitations is strongly influenced by these underlying characteristics.

We studied the impact of charge neutrality and $\beta$ equilibrium on the collective modes of npe and npe$\mu$ matter, with particular emphasis on the coupling between nuclear and plasmon excitations. In charge-neutral, $\beta$-equilibrated matter, the longitudinal response of a relativistic degenerate lepton gas displays two branches for each lepton species: a low-energy sound-like mode and a higher-energy plasmon-like mode, both limited by a cutoff frequency.

The inclusion of leptons significantly enriches the excitation spectrum. Neutron and proton modes in npe and npe$\mu$ matter remain essentially soundlike excitations. In contrast, plasmon-like behavior of the proton mode appears only in np matter when the Coulomb interaction is included, particularly at low momentum transfer \cite{Baldo:2008pb}. In npe matter, the dominant coupling occurs between the electron and proton modes, while in npe$\mu$ matter additional mixing arises through the coupling of the electron and muon modes to the proton mode.

We find that the soundlike lepton branch couples to nuclear modes when the latter possess a predominantly isovector character and the momentum transfer is sufficiently small. For larger momentum transfers or lower proton fractions, the coupling shifts to densities above the isovector–isoscalar crossing density. However, at low momentum transfer and sufficiently large proton fractions, the coupling may already occur at low densities. Consequently, in npe and npe$\mu$ matter, the onset density of the nuclear isovector mode depends sensitively on both the proton fraction and the momentum transfer.

The calculations were performed within the relativistic mean-field (RMF) framework using the NL3, NL3$\omega\rho$, and FSU2H parametrizations in a nonlinear RMF description of nuclear matter. Although the quantitative details are model dependent, the qualitative features discussed here are expected to be robust. Density-dependent RMF models may, nevertheless, exhibit a modified dependence on the momentum transfer.

These results are relevant for astrophysical environments such as neutron stars and core-collapse supernovae, where matter is charge neutral and in $\beta$ equilibrium. In particular, plasmon properties affect neutrino-emission mechanisms, including plasmon decay into neutrino–antineutrino pairs, and therefore may influence transport phenomena in dense matter.

\section*{ACKNOWLEDGMENTS}
The present work was partially supported by Fundação para a Ciência e a Tecnologia (FCT) (PT), under the  project UID/04564/2025 (doi:10.54499/UID/04564/2025). S.S.A. acknowledges partial support from CNPq-308963/2023-7 and INCT-FNA. %
\bibliographystyle{apsrev4-1}
\bibliography{bibliography}

\end{document}